\documentclass[twocolumn]{aastex61}

\usepackage{graphicx}
\usepackage{subfigure}
\usepackage{multirow}
\usepackage{comment}
\usepackage{natbib}
\usepackage{hyperref}
\usepackage{mathtools}
\usepackage{mathrsfs}
\usepackage{fontenc}
\usepackage{color}
\usepackage{url}
\usepackage{hyperref}
\usepackage{gensymb}
\usepackage{pifont}
\usepackage{textcomp}
\usepackage{color}
\newcommand{\OIII}{[O{\sevenrm\,III}]}

\newcommand{\loiii}{$L_{\text{[O \textrm{\tiny III}]}}$}
 
 \font\sevenrm=cmr7 scaled 1000

\def\lax{{$\mathrel{\hbox{\rlap{\hbox{\lower4pt\hbox{$\sim$}}}\hbox{$<$}}}$}}
\def\gax{{$\mathrel{\hbox{\rlap{\hbox{\lower4pt\hbox{$\sim$}}}\hbox{$>$}}}$}}

\begin{document}

\title{The Infrared Emission and Opening Angle of the Torus in Quasars}

\shorttitle{TORUS in QUASARS}

\shortauthors{ZHUANG ET AL.}

\author{Ming-Yang Zhuang}
\email{mingyangzhuang@pku.edu.cn}
\affil{Kavli Institute for Astronomy and Astrophysics, Peking University,
Beijing 100871, China}
\affil{Department of Astronomy, School of Physics, Peking University,
Beijing 100871, China}

\author{Luis C. Ho}
\affil{Kavli Institute for Astronomy and Astrophysics, Peking University,
Beijing 100871, China}
\affil{Department of Astronomy, School of Physics, Peking University,
Beijing 100871, China}

\author{Jinyi Shangguan}
\affil{Kavli Institute for Astronomy and Astrophysics, Peking University,
Beijing 100871, China}
\affil{Department of Astronomy, School of Physics, Peking University,
Beijing 100871, China}

\begin{abstract}
According to the unified model of active galactic nuclei (AGNs), a putative 
dusty torus plays an important role in determining their external appearance.
However, very limited information is known about the physical properties of 
the torus.  We perform detailed decomposition of the infrared 
($1-500\,\mu$m) spectral energy distribution of 76 $z < 0.5$ 
Palomar-Green quasars, combining photometric data from 2MASS, {\it WISE}, and
{\it Herschel} with {\it Spitzer}\ spectroscopy.  Our fits favor recent torus 
spectral models that properly treat the different sublimation temperatures of 
silicates and graphite and consider a polar wind component.  The AGN-heated 
dust emission from the torus contributes a significant fraction ($\sim 70\%$)
of the total infrared ($1-1000\, \mu$m) luminosity.  The torus luminosity correlates 
well with the strength of the ultraviolet/optical continuum and the broad $\rm{H\beta}$ 
emission line, indicating a close link between the central ionization source 
and re-radiation by the torus.  Consistent with the unified model, most 
quasars have tori that are only mildly inclined along the line-of-sight.  The
half-opening angle of the torus, a measure of its covering factor, declines 
with increasing accretion rate until the Eddington ratio reaches $\sim 0.5$, 
above which the trend reverses.  This behavior likely results from the change 
of the geometry of the accretion flow, from a standard geometrically thin disk 
at moderate accretion rates to a slim disk at high accretion rates.
\end{abstract}

\section{Introduction}
\label{sec:intro}

Active galactic nuclei (AGNs) release prodigious amounts of energy from 
accretion of matter by massive black holes (BHs) residing in the center of 
galaxies \citep{1969Natur.223..690L, 1984ARA&A..22..471R}.  Many 
attempts have been made to explain the tremendous observed diversity of 
AGNs through ``unified'' models \citep{1993ARA&A..31..473A, 1995PASP..107..803U, 2015ARA&A..53..365N}. 
In these models, a small-scale (\lax\ 1 pc) dusty torus plays an important 
role in separating type 1 and type 2 AGNs by reprocessing the ultraviolet,
optical, and X-ray radiation into the infrared (IR) band, and by blocking 
photons from the broad-line region from certain viewing angles. 

Early mid-IR observations with the Very Large Telescope Interferometer by 
\citet{2007A&A...474..837T} provided strong evidence supporting the existence of
a clumpy-structured torus \citep[e.g.,][]{1988ApJ...329..702K, 2005A&A...436...47D} 
rather than a smooth torus \citep[e.g.,][]{1992ApJ...401...99P, 1995MNRAS.273..649E}.
Intensive efforts have been invested to model the emission from the torus 
\citep{2006A&A...452..459H, 2008A&A...482...67S, 2008ApJ...685..160N, 
2008ApJ...685..147N, 2010A&A...515A..23H, 2012MNRAS.420.2756S, 
2015A&A...583A.120S}. These models have been extensively employed to 
investigate the torus covering factor and intrinsic differences between type 1 
and type 2 AGNs \citep[e.g.,][]{2011ApJ...736...82A, 2016MNRAS.458.2288S, 
2017MNRAS.472.3492E}.  The covering factor of the torus can also be 
constrained using gas column densities probed by X-ray observations 
\citep[e.g.,][]{2007ApJ...664L..79U, 2012MNRAS.423..702B}. 
\citet{2017Natur.549..488R} report that the torus covering factor is linked 
with the Eddington ratio of the BH. 

Subsequent interferometric observations \citep{2009MNRAS.394.1325R, 
2013ApJ...771...87H, 2014A&A...565A..71L, 2016A&A...591A.128L} reveal that 
the bulk of the mid-IR emission of AGNs actually arises from a polar-extended component, 
which dominates the energy output in that band, while the near-IR emission 
still emanates from a classical, small-scale disk-like component. This discovery 
radically alters the traditional view of a single torus structure and demands 
an update of the present torus models.  Moreover, detailed scrutiny of the
spectral energy distribution (SED) of AGNs \citep{2011ApJ...729..108D, 
2012MNRAS.420..526M} consistently finds that an extra, high-temperature 
blackbody component is needed to account for the near-IR emission in type 1 
AGNs, indicating the existence of hot graphite grains not fully represented 
in current models of clumpy tori \citep[e.g.,][]{2008ApJ...685..147N, 
2008ApJ...685..160N, 2010A&A...515A..23H, 2015A&A...583A.120S}.  
This shortcoming has been addressed recently by \citet{2017MNRAS.470.2578G}, 
who incorporated more physical dust sublimation temperatures for silicates and 
graphite into their torus models.  \citet{2017ApJ...838L..20H} further 
added a polar wind component to mimic the structures seen in the latest 
interferometric observations.  

We apply these newly developed models to investigate the physical 
properties of the torus in a large sample of low-redshift quasars, using the 
comprehensive set of high-quality IR SEDs spanning $\sim 1-500\,\mu$m assembled
by \citet{2018ApJ...854..158S}.  The SEDs combine both photometric data and mid-IR 
spectroscopy.  We use our recently developed Bayesian Markov Chain Monte Carlo 
(MCMC) method to decompose the SEDs into their main constituent components, 
paying special emphasis on evaluating the performance of the latest spectral 
templates for the AGN torus.  We quantify the fractional contribution of the 
torus luminosity to the total IR energy budget and study the inclination 
angle and covering factor of the torus.

This paper is structured as follows. We introduce the torus models used in 
this paper in Section \ref{sec:models}. We show the results of the SED 
fitting in Section \ref{sec:results} and discuss the properties of the torus 
in Section \ref{sec:4}. Conclusions are presented in Section \ref{sec:conclusion}. 
This work adopts the following parameters for a $\Lambda$CDM cosmology: 
$\Omega_m=0.308$, $\Omega_{\Lambda}=0.692$, and $H_0 = 67.8\, \mathrm
{km\,s^{-1}\,Mpc^{-1}}$ \citep{2016A&A...594A..13P}.

\begin{figure*}[!ht]
\centering
\includegraphics[width=\textwidth]{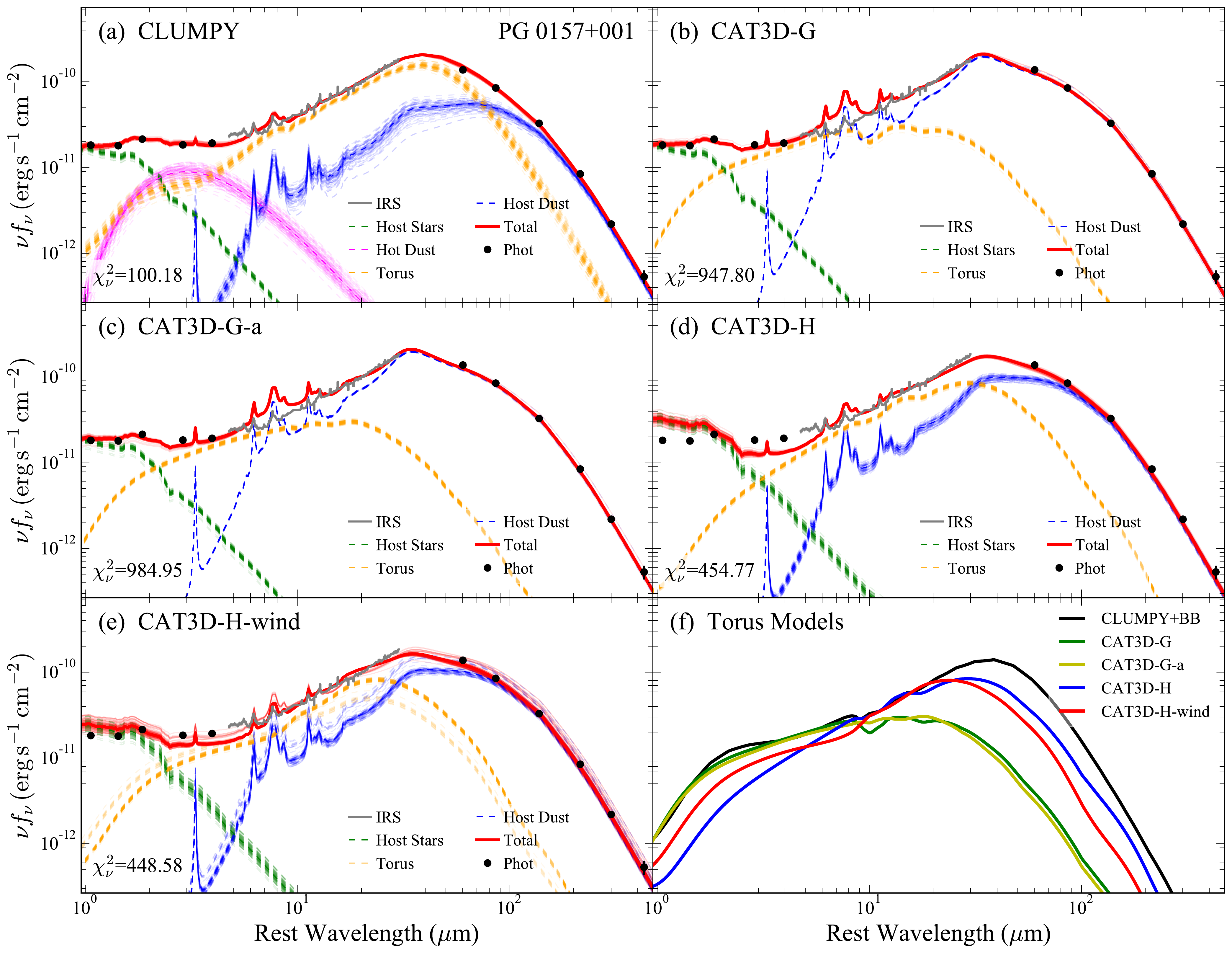}
\caption{Examples of SED fitting for PG~0157+001, employing torus models from
(a) CLUMPY, (b) CAT3D-G, (c) CAT3D-G-a, (d) CAT3D-H, and (e) CAT3D-H-wind.
Panel (f) compares the torus components of all five torus models.  In panels
(a)--(e), the grey line represents the {\it Spitzer}/IRS spectrum, and the
black points show photometric data from 2MASS, {\it WISE}, and {\it Herschel}.
The dashed lines are the best-fit models: host galaxy stars (BC03; green), torus (orange), 
and host galaxy dust (DL07; blue). 
In the case of panel (a), the CLUMPY torus model is supplemented with an additional 
blackbody component for very hot dust (magenta). 
The combined, best-fit total 
model is the red solid line. To visualize the model uncertainties, the associated thin lines in
light color represent 100 sets of models with parameters drawn randomly from the space 
sampled by the MCMC algorithm. 
The reduced chi-squared ($\chi_{\nu}^{2}$) for the part of the SED covering the IRS 
spectrum is shown on the lower-left corner of panels (a)--(e) to quantify the goodness-of-fit 
(see Appendix \ref{app:B} for details).
The complete figure set (76 figures) of fitting results with CAT3D-H-wind torus model is 
available in the online journal.}
\label{fig:1}
\end{figure*}

\begin{figure*}[!ht]
\centering
\includegraphics[width=\textwidth]{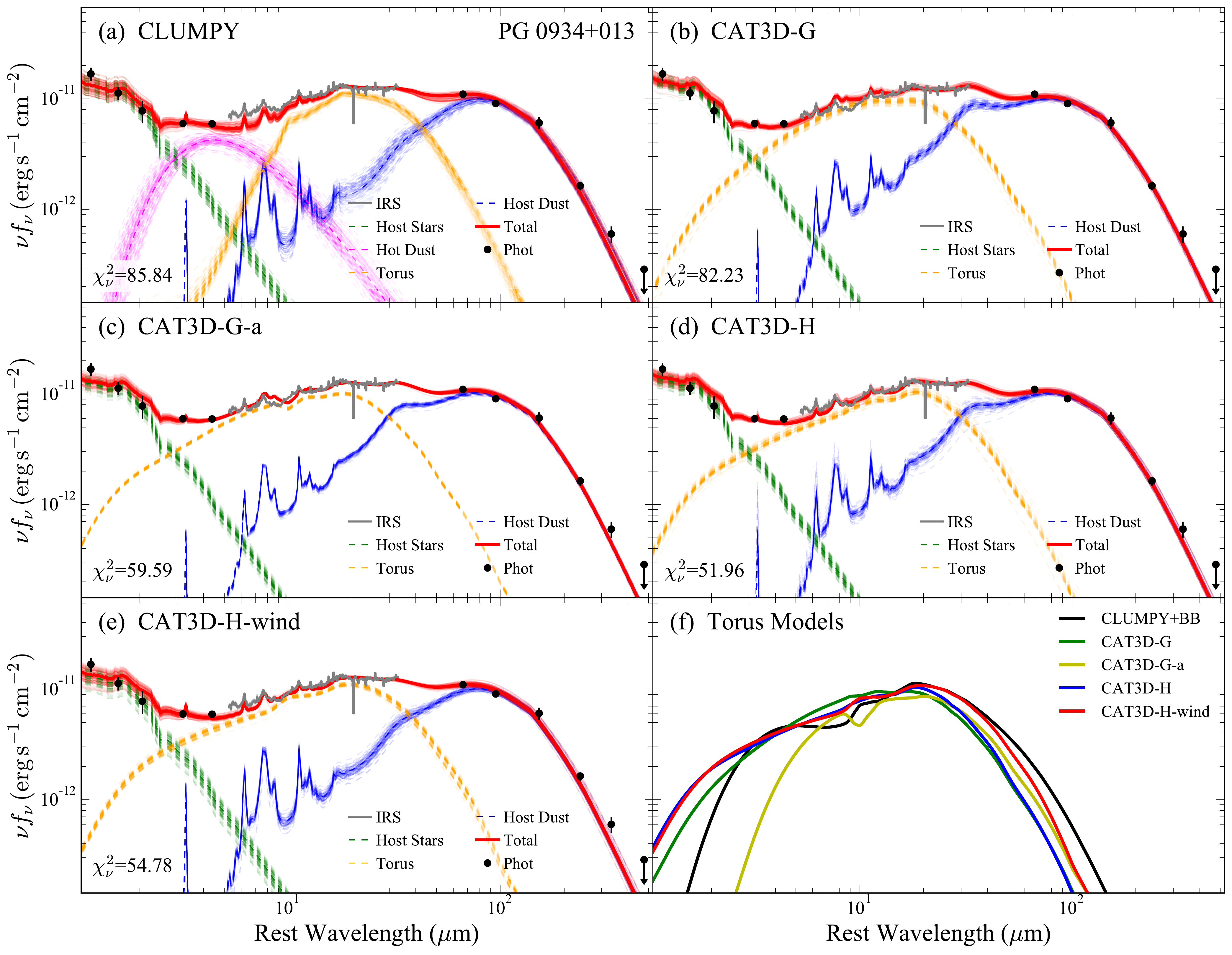}
\caption{Same as Figure \ref{fig:1}, but for PG~0934+013.}
\label{fig:2}
\end{figure*}

\begin{figure*}[!ht]
\centering
\includegraphics[width=\textwidth]{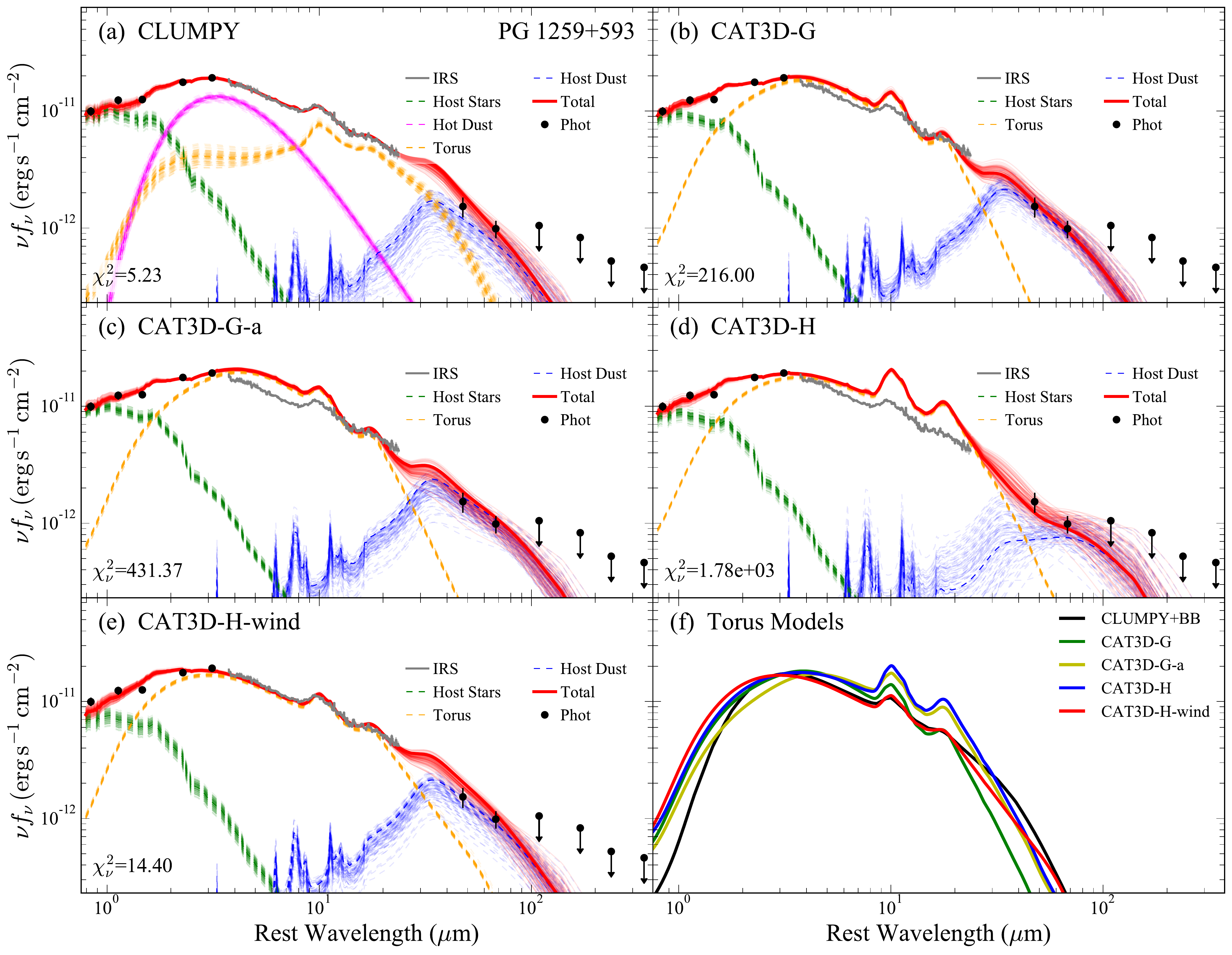}
\caption{Same as Figure \ref{fig:1}, but for PG~1259+593.}
\label{fig:3}
\end{figure*}

\section{models}
\label{sec:models}

We use up to four components to fit the IR SED (Figure \ref{fig:1}): (1) host 
galaxy stellar emission peaking in the near-IR, (2) AGN torus emission peaking 
in the near-IR and mid-IR, (3) cold dust emission from the large-scale 
interstellar medium of the host galaxy peaking in the far-IR, and, if necessary,
(4) an extra synchrotron jet component in the case of radio-loud objects. We 
adopt the same models as \citet{2018ApJ...854..158S} for the stellar emission 
(\citealp{2003MNRAS.344.1000B}; BC03), interstellar dust emission 
(\citealp{2007ApJ...657..810D}; DL07), and synchrotron radiation 
(\citealt{2014SSRv..183..371P}; broken power-law).  For the torus emission, 
apart from the CLUMPY model \citep{2008ApJ...685..147N, 2008ApJ...685..160N} 
and the complementary blackbody component (BB) added to account for emission 
from very hot dust as employed by \citet{2018ApJ...854..158S}, here we make use of 
two sets of newly calculated torus models (``Clumpy AGN Tori in a 3D geometry",
CAT3D, \citealp{2010A&A...515A..23H}) recently developed by 
\citet{2017MNRAS.470.2578G} and \citet{2017ApJ...838L..20H}.

\citet{2008ApJ...685..147N, 2008ApJ...685..160N} developed a formalism to perform radiative 
transfer calculations of clumpy clouds in a torus, which enables a large range 
of dust temperatures to coexist at the same distance from the central radiation 
source. They assume a sublimation temperature of $\sim 1500$ K for both 
silicate and graphite dust. Their model has seven free parameters to 
characterize the properties of the torus: (1) the optical depth  $\tau_V$ of 
individual clouds, (2) the power-law index $q$ of the radial distribution of 
clouds, (3) the ratio $Y$ between the outer and inner sublimation radius 
$r_{\mathrm{sub}}$, (4) the average number of clouds in the equatorial 
direction $N_0$, (5) the standard deviation $\sigma$ of the Gaussian 
distribution of the number of clouds in the vertical direction, (6) the 
observer's viewing angle $i$ with respect to the normal of the torus plane, 
and (7) a normalization factor $L$. Together with two additional free 
parameters for the BB to account for the very hot dust component, there are a 
total of nine free parameters. The CLUMPY model has more than 1.2 million 
spectral templates, covering a large range of parameter space.

Building upon the original framework of the torus model of \citet{2010A&A...515A..23H}, 
\citet{2017ApJ...838L..20H} expanded the CAT3D model to account 
for the mid-IR observational evidence of extended dust emission emanating from 
the polar direction of the nuclear regions of AGNs \citep{2009MNRAS.394.1325R, 
2013ApJ...771...87H, 2014A&A...565A..71L, 2016A&A...591A.128L}, and to allow 
for the possibility that silicates and graphite have different sublimation 
temperatures and grain size distributions.  While silicates sublimate at 
temperatures higher than $\sim 1200$ K, graphite can withstand temperatures up 
to $\sim 1900$ K \citep[e.g.,][]{1987ApJ...320..537B, 2007A&A...476..713K, 
2012MNRAS.420..526M, 2017ApJ...838L..20H}.
Supposing that the extended nuclear dust arises from some kind of outflow, 
they consider a wind component with the shape of a hollow cone in the polar 
region of the AGN, possibly formed by dust clouds lifted by radiation pressure 
near the dust sublimation radius. Three free parameters specify the properties 
of the torus properties: the power-law index $a$ of the cloud radial 
distribution of the form $r^a$, with $r$ the distance from the center in 
units of $r_{\rm sub}$; the dimensionless scale height $h$ of the Gaussian 
distribution of clouds in the vertical direction of the form $\exp\{-z^2/2(hr)^2\}$, with $z$ the vertical distance 
distribution from the mid-plane; and the average number $N_0$ of clouds along the equatorial 
line-of-sight.  The wind itself is characterized by five free parameters:
the radial distribution $a_w$ of dust clouds, the half-opening angle 
$\theta_w$, the angular width $\sigma_{\theta}$, and a wind-to-disk ratio 
$f_{\rm wd}$, which defines the ratio between the number of clouds along the 
cone and $N_0$. Together with the inclination (i.e. viewing) angle $i$ and 
normalization factor $\log L$, there are nine free parameters in total.
The model holds constant three additional parameters, namely the outer radius 
of the torus and wind $R_{\rm out}$, the size of each cloud $R_{\rm cl}$, and 
the optical depth of each cloud $\tau_V$. Two sets of torus models are 
provided, with (CAT3D-H-wind) and without (CAT3D-H) winds.  Both cases properly
treat the size distribution and dust sublimation temperature for silicates 
and graphite grains.
The parameter space for the two sets of models is different, and the 
number of spectral templates is also different. \citet{2017ApJ...838L..20H} 
provide 132,300 templates for CAT3D-H-wind and 1,078 templates for CAT3D-H.

Similar to \citet{2017ApJ...838L..20H}, \citet{2017MNRAS.470.2578G}
improved the original CAT3D model with a more realistic physical treatment of 
differential dust grain sublimation and anisotropic AGN emission, motivated by
the expectation that the ultraviolet photons produced by the accretion disk are 
angularly dependent ($\cos i$).  Again, two sets of torus models are available.
One considers only the effects of different sublimation temperatures (CAT3D-G),
and the other includes, in addition, anisotropic AGN emission (CAT3D-G-a). 
Both sets of models have the same free parameters but cover a different range 
of values: power-law index of cloud radial distribution $a$, half-opening 
angle $\theta_0$, number of clouds along equatorial direction $N_0$, 
inclination angle $i$, and normalization factor $L$.  The optical depth 
$\tau_V$ and outer radius of the torus $R_{\rm out}$ are kept fixed.  The CAT3D-G 
model contains 1,232 templates, each with 10 random distributions of clouds;
the CAT3D-G-a model only covers 427 templates with each having 20 random 
distributions of clouds \citep{2017MNRAS.470.2578G}.

\begin{figure}[t]
\centering
\includegraphics[width=0.47\textwidth]{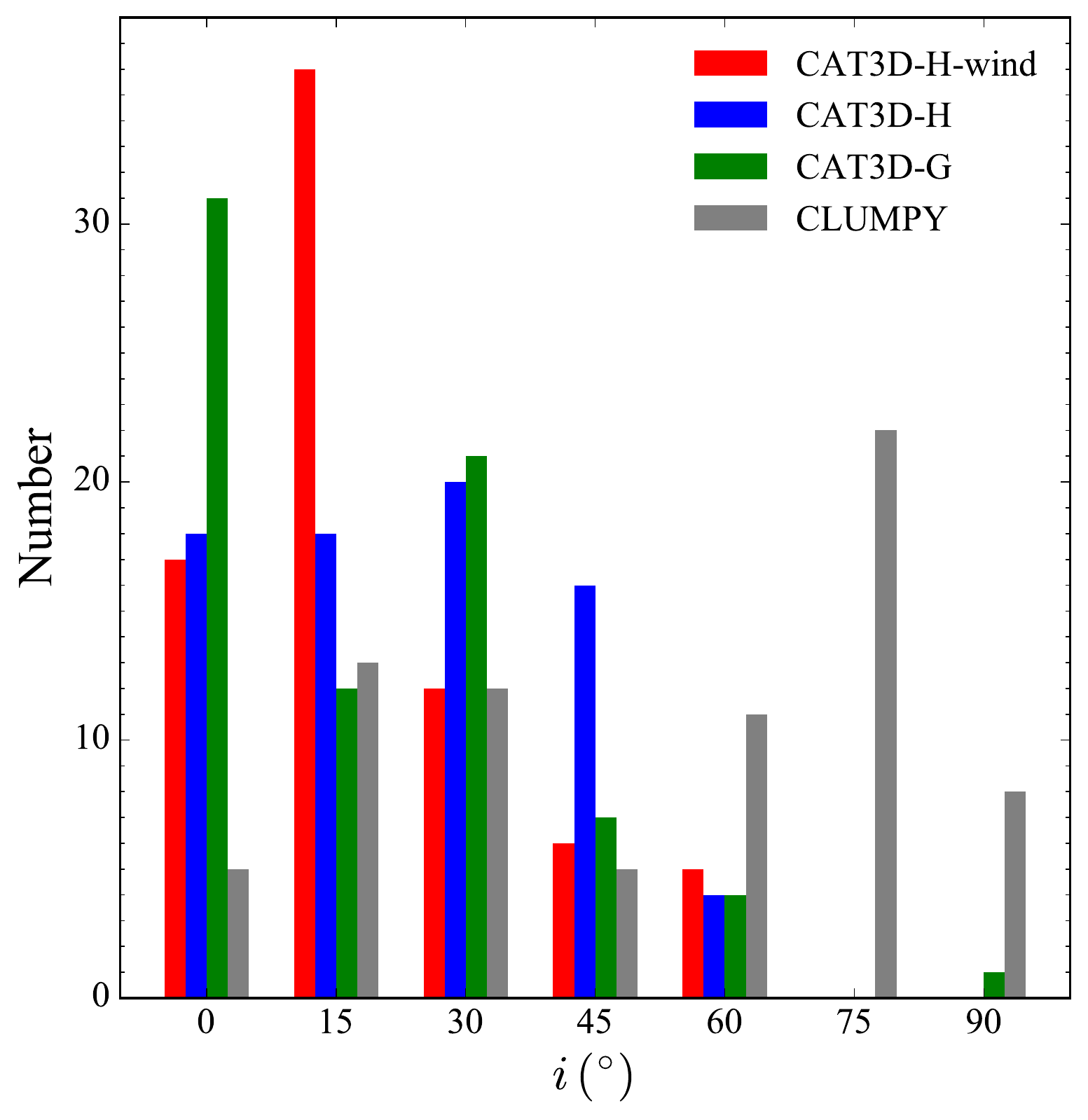}
\caption{Inclination angle ($i$) of the torus derived from different models:
CAT3D-H-wind (red), CAT3D-H (blue), CAT3D-G (green), and CLUMPY (gray). The 
distributions of $i$ derived from the three CAT3D models are similar
and favor low values.  By contrast, the CLUMPY model gives a much broader
distribution of $i$, with a significant fraction of $i$ larger than 60\degree.}
\label{fig:4}
\end{figure}

\begin{figure}[h]
\centering
\includegraphics[width=0.47\textwidth]{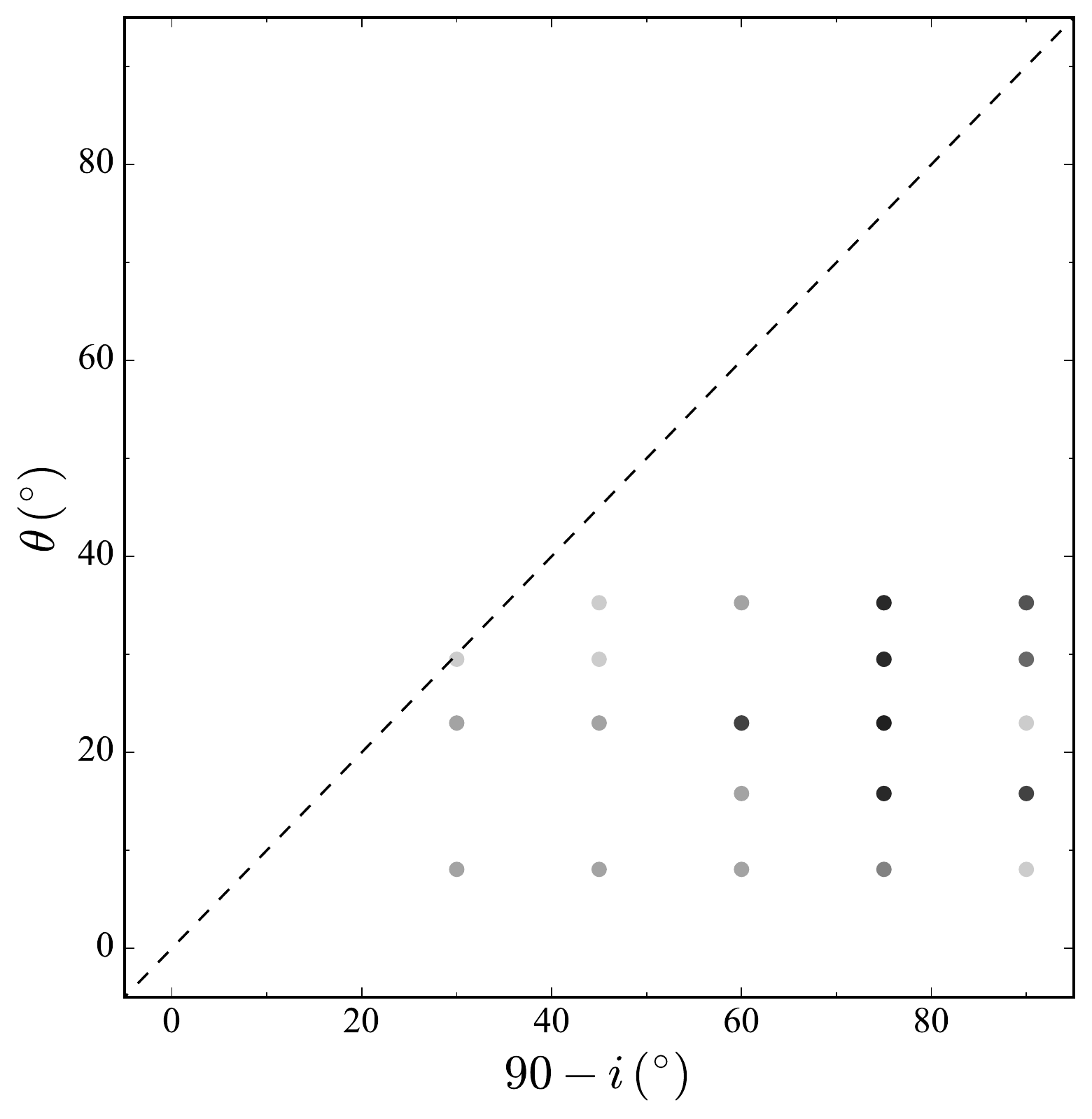}
\caption{Comparison of the torus inclination angle $i$ (complementary angle
$90\degree-i$) and half-opening angle $\theta$ from the CAT3D-H-wind model.  The
dashed line is the 1:1 relation.  Because of the discreteness of the parameter
space, we do not show error bars. Darker points indicate more objects.}
\label{fig:5}
\end{figure}

\section{Results}
\label{sec:results}

Our analysis is based on the database of \citet{2018ApJ...854..158S}, who presented 
complete IR ($\sim 1-500\,\mu$m) SEDs of a sample of 87 low-redshift
($z < 0.5$) type 1 (broad-lined) quasars selected by \citet{1992ApJS...80..109B} 
from the Palomar-Green (PG) survey \citep{1983ApJ...269..352S}. The SEDs were 
assembled using photometric data acquired from the Two Micron All-Sky Survey 
(2MASS), {\it Wide-field Infrared Survey Explorer}\ ({\it WISE}), and 
{\it Herschel Space Observatory}\ ({\it Herschel}), in concert with 
low-resolution mid-IR spectra taken with the Infrared Spectrometer (IRS) on 
the {\it Spitzer Space Telescope}\ ({\it Spitzer}). Excluding 11 objects with 
insufficient far-IR detections from {\it Herschel}, our final sample consists 
of 76 PG quasars.

We fit the SEDs using the method newly developed by \citet{2018ApJ...854..158S}.
Physical models of host galaxy starlight, AGN torus, and galaxy-scale cold dust 
are combined to fit the integrated quasar SED using an MCMC method.  The 
photometric and spectroscopic data are fit simultaneously, incorporating both 
detections and upper limits.  The mid-IR photometry, being redundant with the 
IRS spectra, has a negligible effect on the likelihood; we mitigate this effect
by modeling the covariance of the residuals between the spectrum and the model.
Five sets of models are applied, corresponding to each of the five different 
sets of torus templates (CLUMPY+BB and the four versions of CAT3D).  For the
models of \citet{2017MNRAS.470.2578G}, we use the first of 10 random sets of 
CAT3D-G templates, and the median value of 20 random sets of CAT3D-G-a 
templates as the final template for each configuration.  Following 
\citet{2018ApJ...854..158S}, we choose a 5 Gyr stellar population with a 
\citet{2003PASP..115..763C} initial mass function from BC03 for the starlight
component, and we employ emission templates from DL07 for the host galaxy 
dust component\footnote{For completeness, we note that the choice of torus 
model adopted for the SED fit makes very little difference on the derived dust 
masses.  This is quantified in Appendix \ref{app:A}.}. 
For 11 radio-loud objects, we incorporate a synchrotron emission component for
the jet, fitting the SED with archival radio measurements collected in 
\citet{2018ApJ...854..158S}.  However, the synchrotron emission is not dominant at 
sub-millimeter wavelengths for any of the objects.

\subsection{Fitting Results}
\label{sec:3.1}

Among the five torus models, CAT3D-H-wind, which incorporates a polar wind 
component, and CLUMPY provide excellent overall fits to almost all the objects. 
The other three CAT3D models produce good fits for less than half of the sample.
A quantitative assessment is given in Appendix \ref{app:B}.
Three examples are given in Figures \ref{fig:1}--\ref{fig:3}, which show fits 
using all five torus models, together with a direct comparison of the best-fit 
torus components.  In the case of PG~1259+593 (Figure \ref{fig:3}), which has 
prominent silicate emission at $\sim 10$ $\mu$m and very strong hot dust 
emission at $\sim 5$ $\mu$m,  the new torus models that lack a wind component 
(CAT3D-G, CAT3D-G-a, CAT3D-H) clearly perform poorly. 
Without an additional polar wind component, these three models always predict 
much stronger silicate emission. This is likely because for torus models with 
only a toroidal structure, the distribution of silicates is directly tied to 
that of graphite, such that stronger hot dust emission always leads to 
stronger silicate emission (see Appendix \ref{app:B} for details).  By 
contrast, the CAT3D-H-wind model matches closely the overall SED.  This 
strongly suggests that PG~1259+593 indeed has a dusty polar wind.  As discussed
in \citet{2017ApJ...838L..20H}, the wind component dominates the mid-IR 
emission while the torus itself is responsible for the hot near-IR emission 
\citep{2013ApJ...771...87H}.  The near-IR emission and the bulk of the mid-IR 
emission are isolated naturally, obviating the need to have such strong 
silicate emission.  
Not surprisingly, the CLUMPY model, when combined with an extra 
hot BB component, has the flexibility to give an equally good fit.  However, 
the ad hoc nature of the hot BB component renders this option less desirable.
The CAT3D-G-a model generally performs most poorly, possibly because of the 
limited number of available spectral templates; we do not consider this 
model further in the following analysis.

\subsection{Inclination Angle $i$ and Half-opening Angle $\theta$}
\label{sec:3.2}

As our sample consists of type 1 quasars, their broad-line region is directly 
visible to us. This generally restricts the inclination angle along the 
line-of-sight $i$ to be relatively low.  Furthermore, the torus should not 
block the photons from the broad-line region, which means that the 
complementary angle of the inclination ($90\degree - i$) should be larger than the 
half-opening angle $\theta$, although there is still a chance for us to see 
the broad-line region at large inclination because of the clumpy structure of 
the torus \citep{2008ApJ...685..147N}. According to the definition of the 
scale height $h$ for the CAT3D-H and CAT3D-H-wind scenarios, $\theta = 
\sqrt{2}\arctan h$.

Figure \ref{fig:4} shows the values of $i$ derived from the four torus models. 
All of the three variants of the CAT3D templates deliver inclination angles clustered  
toward relatively low values (mostly $i$ \lax\ $45\degree$), consistent with 
expectations.  By contrast, the CLUMPY (plus BB) model, despite its success 
in reproducing the overall SED, yields a very broad distribution of 
inclinations, with more than half of the sample having $i$ \gax\ $60\degree$. 
Such large inclinations are inconsistent with the type 1 nature of these 
sources.  Apart from the unrealistically large values of $i$, the CLUMPY model,
as previously mentioned, needs to be supplemented with an extra, artificially 
added blackbody component to compensate for the lack of emission from hot dust 
at $\sim 5$ $\mu$m \citep{2011ApJ...729..108D, 2012MNRAS.420..526M}. This 
would result in a misleading goodness-of-fit achieved by CLUMPY. Therefore, 
previous torus parameters 
derived from CLUMPY \citep[e.g.,][]{2009ApJ...707.1550N, 2017MNRAS.464.2139A} 
should be treated with caution. 
The tendency for CLUMPY+BB to yield large inclinations can be understood.
The BB component usually occupies as much near-IR emission as it can. The CLUMPY 
component is then biased  toward longer wavelengths to compensate, which results in cooler 
temperatures and hence larger inferred inclinations (to block the inner hot dust emission).

In light of these factors, we henceforth only focus on parameters derived 
from the CAT3D-H-wind model, which is not only the most comprehensive and 
most physical, but, as discussed in Section \ref{sec:3.1}, also can 
better capture the full complexities of the observed SEDs.  Figure \ref{fig:5} 
examines the relation between the complementary angle of $i$ and $\theta$.
All the values are located on the lower-right region of the plot (i.e. 
$\theta \ge 90\degree - i$), as expected for type 1 AGNs that are observed 
almost face-on.  This further reinforces our confidence in the physical 
robustness of the CAT3D-H-wind model.  The properties of the torus 
for our sample are given in Table \ref{table:1}.

\begin{deluxetable*}{lRRRRRRRRCCCC}
\tablecaption{Best-fit Parameters for the Torus Model\label{table:1}}
\tablehead{
\colhead{Name} & \colhead{$a$} & \colhead{$h$}& \colhead{$N_0$}& \colhead{$i$}& \colhead{$f_{\rm wd}$}& \colhead{$a_w$}& \colhead{$\theta_w$}& \colhead{$\theta_{\sigma}$}& \colhead{$\log\ L$} & \colhead{$\log L_{\mathrm{total}}$}& \colhead{$\log L_{\mathrm{torus}}$}& \colhead{$\log M_d$}\\
\nocolhead{Name} & \nocolhead{$a$} & \nocolhead{$h$}& \nocolhead{$N_0$}& \colhead{$(\degree)$}& \nocolhead{$f_{\rm wd}$}& \nocolhead{$a_w$}& \colhead{$(\degree)$}& \nocolhead{$\theta_{\sigma}$}& \nocolhead{$\log L$}& \colhead{$(\mathrm{erg\ s^{-1}})$}& \colhead{$(\mathrm{erg\ s^{-1}})$}& \colhead{$(M_{\odot})$}\\
\colhead{(1)} & \colhead{(2)} & \colhead{(3)}& \colhead{(4)}& \colhead{(5)}& \colhead{(6)}& \colhead{(7)}& \colhead{(8)}& \colhead{(9)}& \colhead{(10)}& \colhead{(11)}& \colhead{(12)}& \colhead{(13)}
}
\startdata
 PG $0003+199$  &  $-3.0$  &  $0.3$  &  $ 5.0$  &  $60$  &  $ 0.3$  &  $-1.0$  &  $45$  &  $ 7.0$  &  $43.47_{-0.01}^{+0.00}$  &  $44.49_{-0.00}^{+0.00}$  &  $44.38_{-0.00}^{+0.00}$  &  $6.32_{-0.03}^{+0.03}$ \\
 PG $0007+106$  &  $-3.0$  &  $0.2$  &  $10.0$  &  $30$  &  $0.75$  &  $-1.0$  &  $45$  &  $ 7.0$  &  $43.84_{-0.00}^{+0.01}$  &  $45.23_{-0.01}^{+0.01}$  &  $45.01_{-0.01}^{+0.01}$  &  $7.55_{-0.05}^{+0.04}$ \\
 PG $0026+129$  &  $-3.0$  &  $0.5$  &  $ 5.0$  &  $15$  &  $0.45$  &  $-1.0$  &  $45$  &  $10.0$  &  $44.04_{-0.01}^{+0.01}$  &  $45.31_{-0.01}^{+0.01}$  &  $45.14_{-0.01}^{+0.01}$  &  $6.97_{-0.22}^{+0.21}$ \\
 PG $0049+171$  &  $-3.0$  &  $0.5$  &  $ 5.0$  &  $ 0$  &  $0.75$  &  $-1.5$  &  $45$  &  $ 7.0$  &  $42.96_{-0.01}^{+0.01}$  &  $44.22_{-0.01}^{+0.00}$  &  $44.07_{-0.00}^{+0.01}$  &  $6.30_{-0.18}^{+0.25}$ \\
 PG $0050+124$  &  $-2.0$  &  $0.4$  &  $10.0$  &  $ 0$  &  $0.75$  &  $-0.5$  &  $45$  &  $10.0$  &  $43.94_{-0.00}^{+0.01}$  &  $45.64_{-0.00}^{+0.00}$  &  $45.41_{-0.00}^{+0.01}$  &  $8.21_{-0.01}^{+0.01}$ \\
 PG $0052+251$  &  $-3.0$  &  $0.3$  &  $ 7.5$  &  $60$  &  $0.45$  &  $-1.0$  &  $45$  &  $ 7.0$  &  $44.31_{-0.01}^{+0.01}$  &  $45.48_{-0.01}^{+0.00}$  &  $45.33_{-0.01}^{+0.01}$  &  $8.32_{-0.02}^{+0.04}$ \\
 PG $0157+001$  &  $-1.5$  &  $0.5$  &  $10.0$  &  $30$  &  $0.75$  &  $-1.0$  &  $45$  &  $15.0$  &  $44.62_{-0.02}^{+0.02}$  &  $46.27_{-0.01}^{+0.01}$  &  $45.87_{-0.02}^{+0.02}$  &  $8.63_{-0.03}^{+0.04}$ \\
 PG $0804+761$  &  $-3.0$  &  $0.3$  &  $ 7.5$  &  $15$  &  $ 1.0$  &  $-0.5$  &  $30$  &  $ 7.5$  &  $44.44_{-0.01}^{+0.02}$  &  $45.58_{-0.01}^{+0.01}$  &  $45.52_{-0.01}^{+0.01}$  &  $7.04_{-0.17}^{+0.17}$ \\
 PG $0838+770$  &  $-3.0$  &  $0.2$  &  $10.0$  &  $15$  &  $0.75$  &  $-1.5$  &  $45$  &  $15.0$  &  $43.87_{-0.01}^{+0.01}$  &  $45.26_{-0.00}^{+0.00}$  &  $45.01_{-0.01}^{+0.01}$  &  $8.18_{-0.03}^{+0.03}$ \\
 PG $0844+349$  &  $-2.5$  &  $0.2$  &  $10.0$  &  $15$  &  $0.75$  &  $-1.0$  &  $45$  &  $10.0$  &  $43.31_{-0.00}^{+0.01}$  &  $44.78_{-0.00}^{+0.01}$  &  $44.63_{-0.00}^{+0.00}$  &  $8.00_{-0.05}^{+0.04}$ \\
\enddata
\tablecomments{(1) Object name. (2) Power-law index. (3) Dimensionless scale height of Gaussian distribution for vertical distribution of clouds. (4) Average number of clouds along an equatorial line-of-sight. (5) Inclination. (6) Wind-to-disk ratio. (7) Radial distribution of dust clouds in the wind. (8) Half-opening angle of the wind. (9) Angular width of wind. (10) Luminosity normalization factor. (11) IR luminosity from 1 to 1000 $\mu$m. (12) Torus luminosity from 1 to 1000 $\mu$m. (13) Cold dust mass derived from DL07 model. Upper and lower values represent 84th and 16th percentile of the marginalized poster probability density function. Table 1 is published in its entirety in the machine-readable format. A portion is shown here for guidance regarding its form and content.}
\end{deluxetable*} 

\begin{figure}[t]
\centering
\includegraphics[width=0.47\textwidth]{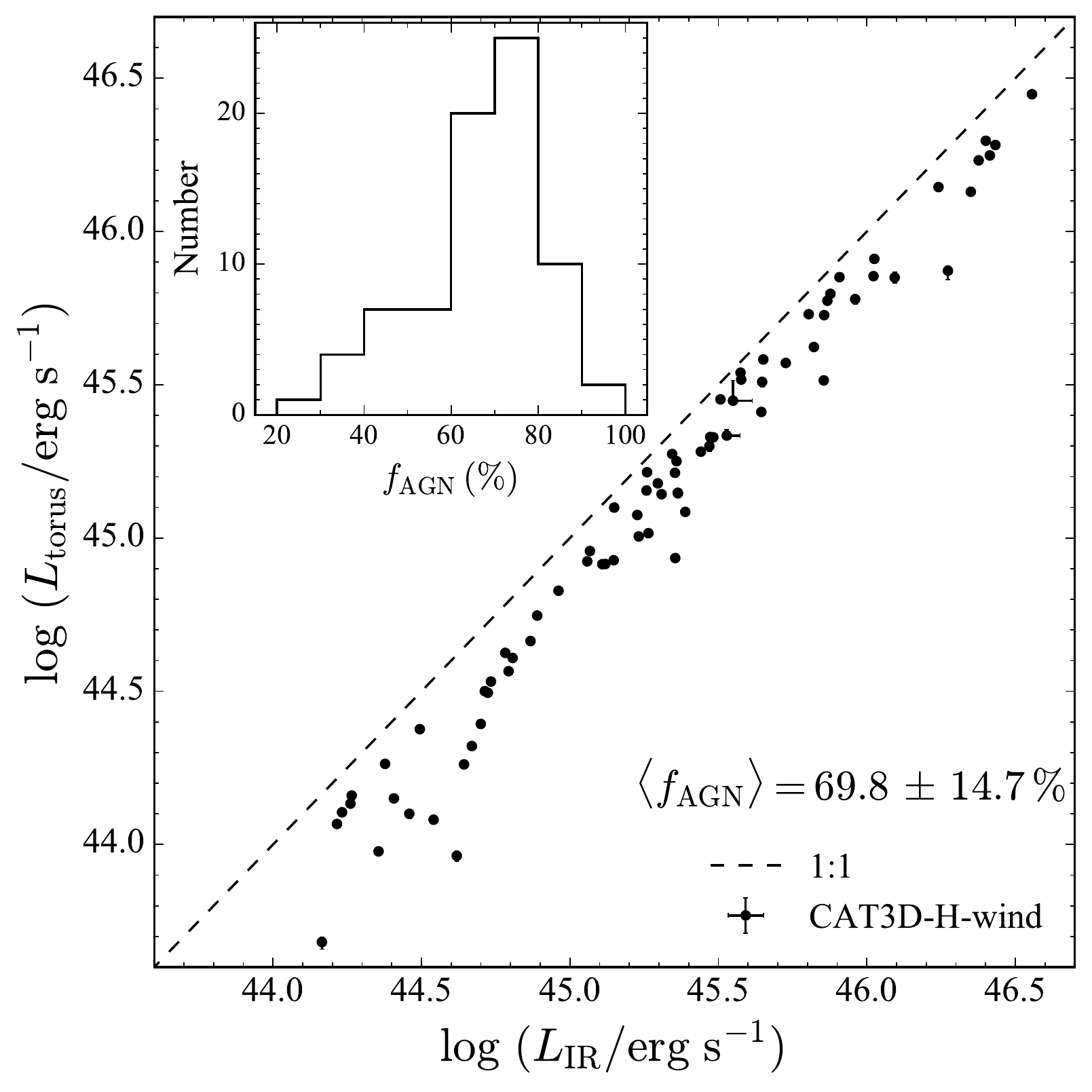}
\caption{Relation between total IR (1--1000 $\mu$m) luminosity
($L_{\mathrm{IR}}$) and the luminosity of the torus ($L_{\rm torus}$) derived
from fits using the CAT3D-H-wind model.  Error bars represents the 68\%
confidence interval determined from the 16th and 84th percentile of the
marginalized posterior probability density function; most error bars are
smaller than the size of the symbols.  The median luminosity fraction of the
torus $f_{\mathrm{AGN}} = 69.8\% \pm 14.7\%$ ($-0.156 \pm 0.113$ dex), as
shown in the inset histogram.}
\label{fig:6}
\end{figure}

\begin{figure*}[t]
\centering
\includegraphics[height=0.7\textheight]{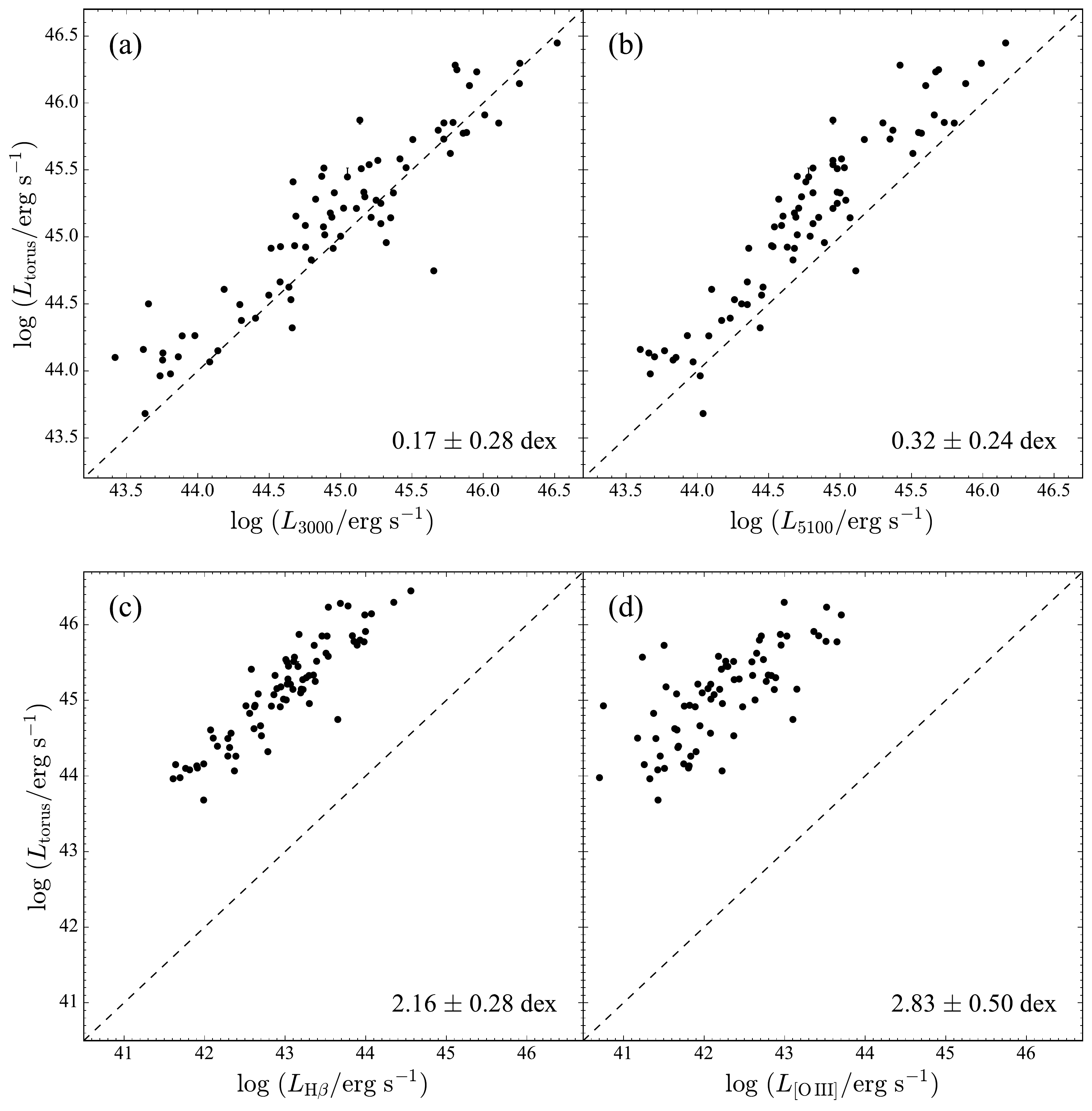}
\caption{Correlation between $L_{\rm torus}$, the luminosity of the torus 
derived from the CAT3D-H-wind model and (a) $L_{3000}$, the monochromatic 
luminosity at 3000 \AA, (b) $L_{5100}$, the monochromatic luminosity at 5100 
\AA, (c) $L_{\mathrm{H\beta}}$, the luminosity of broad H$\beta$, and (d) 
\loiii, the luminosity of \OIII\ $\lambda$5007. 
Median and standard deviation are shown on the bottom-right corner of each panel, 
and the dashed line indicates a one-to-one correlation.
}
\label{fig:7}
\end{figure*}

\begin{figure*}[t]
\centering
\includegraphics[width=\textwidth]{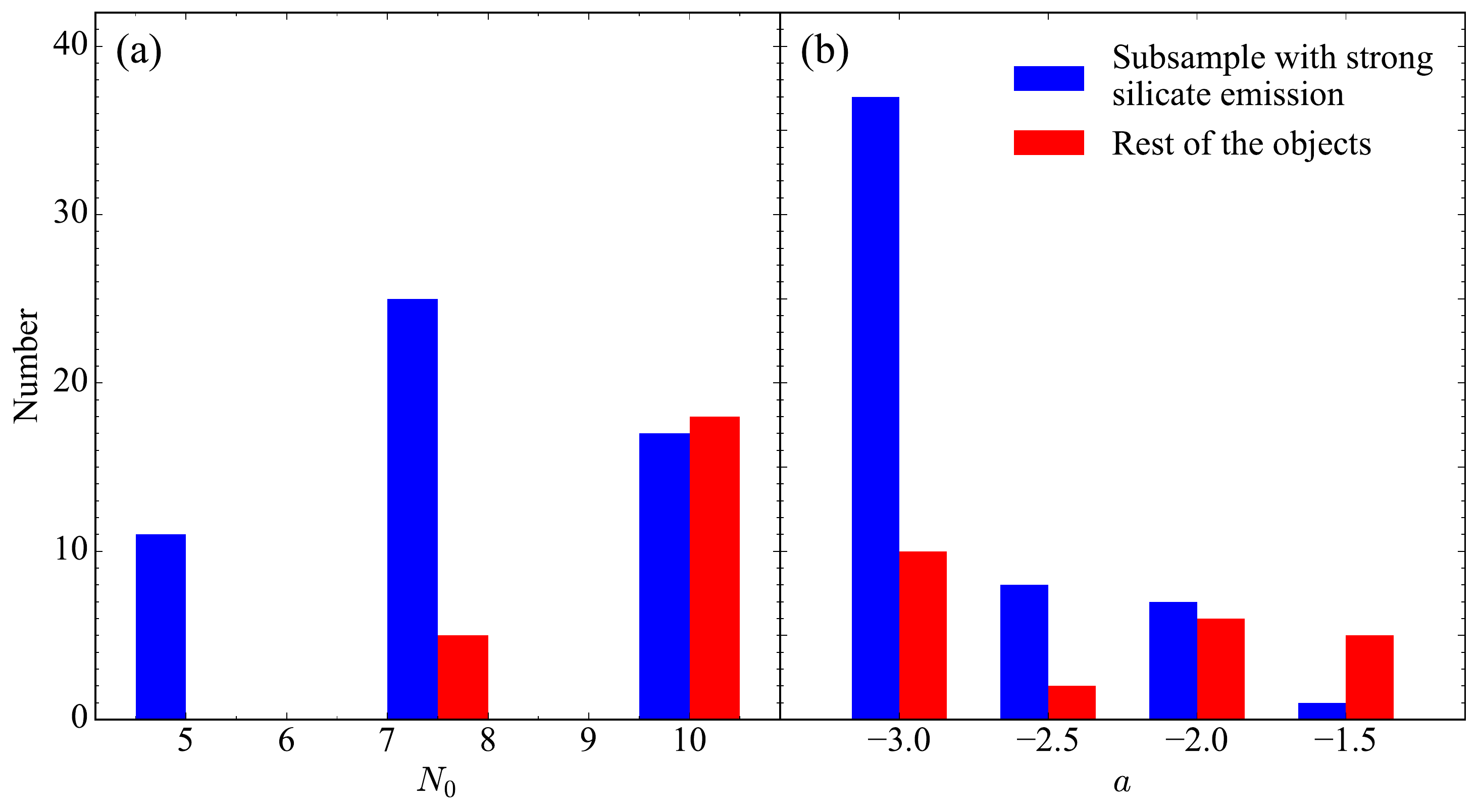}
\caption{
Comparison of the distributions of (a) number of clouds along the equatorial 
line-of-sight ($N_0$) and (b) power-law index of radial density profile ($a$)
for the AGN-dominated subsample (blue) with strong silicate emission 
[$\nu f_{\nu} (9.7 \mu \mathrm{m})/\nu f_{\nu}(14 \mu \mathrm{m})>1$ 
and $f_{\mathrm{AGN}}>50\%$] and the rest of the objects (red). 
}
\label{fig:8}
\end{figure*}

\begin{figure*}[t]
\centering
\includegraphics[width=\textwidth]{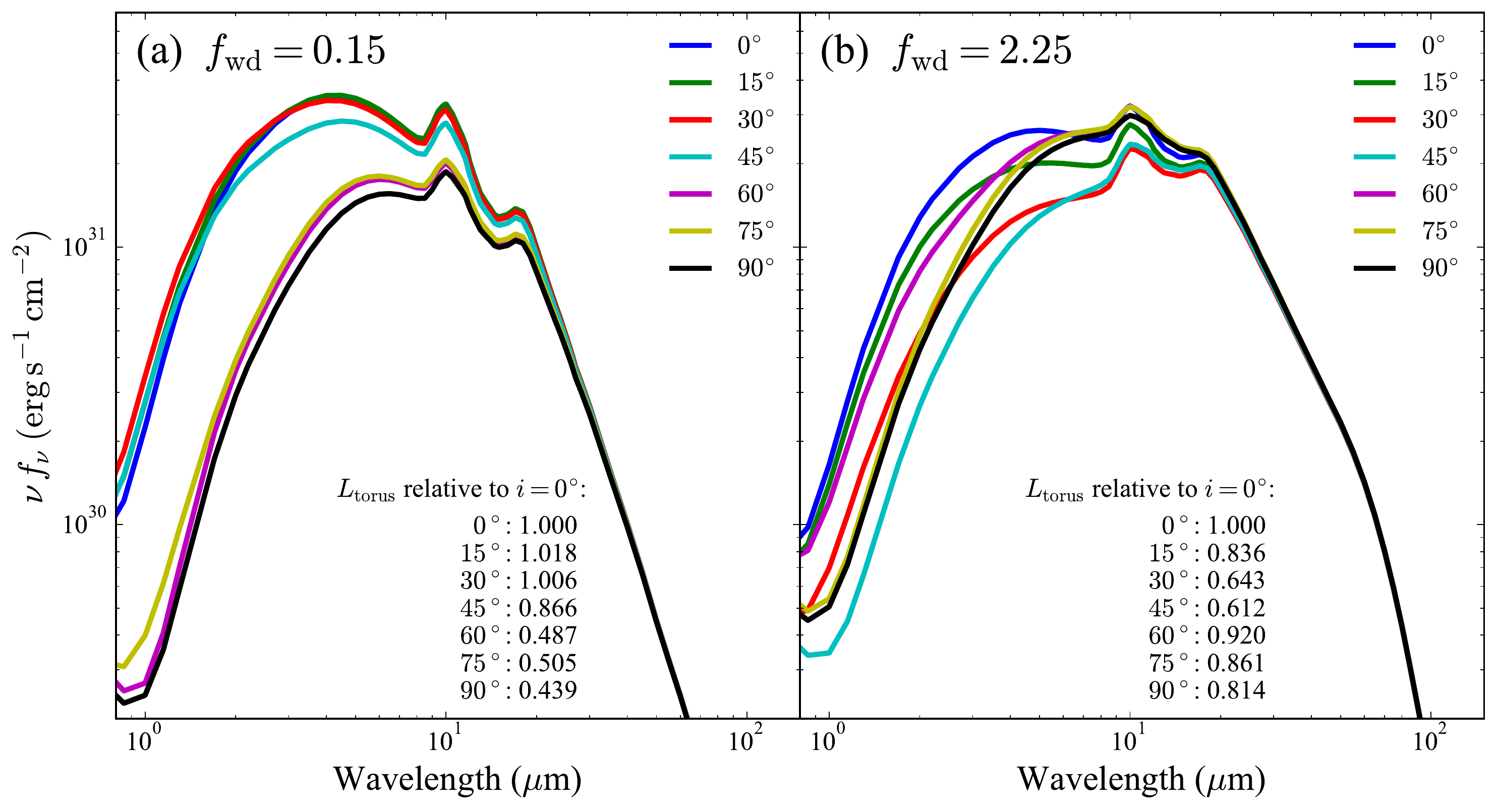}
\caption{The influence of the inclination angle $i$ (0$\degree$ to 90$\degree$)
on the torus SED, for the CAT3D-H-wind model computed with (a) $f_{\rm wd} =
0.15$ and (b) $f_{\rm wd} = 2.25$, with all other model parameters fixed
($N=5$, $a=-3.00$, $h=0.50$, $a_w=-2.00$, $\theta_w=30\degree$, and
$\theta_{\sigma}=15.00$).  The fluxes are scaled to a sublimation
radius of 0.29 pc.  The legend on the bottom part shows the IR
($1 - 1000$ $\mu$m) luminosity of the torus for each specific $i$, relative to
$i = 0\degree$.}
\label{fig:9}
\end{figure*}

\begin{figure*}[t]
\centering
\includegraphics[width=\textwidth]{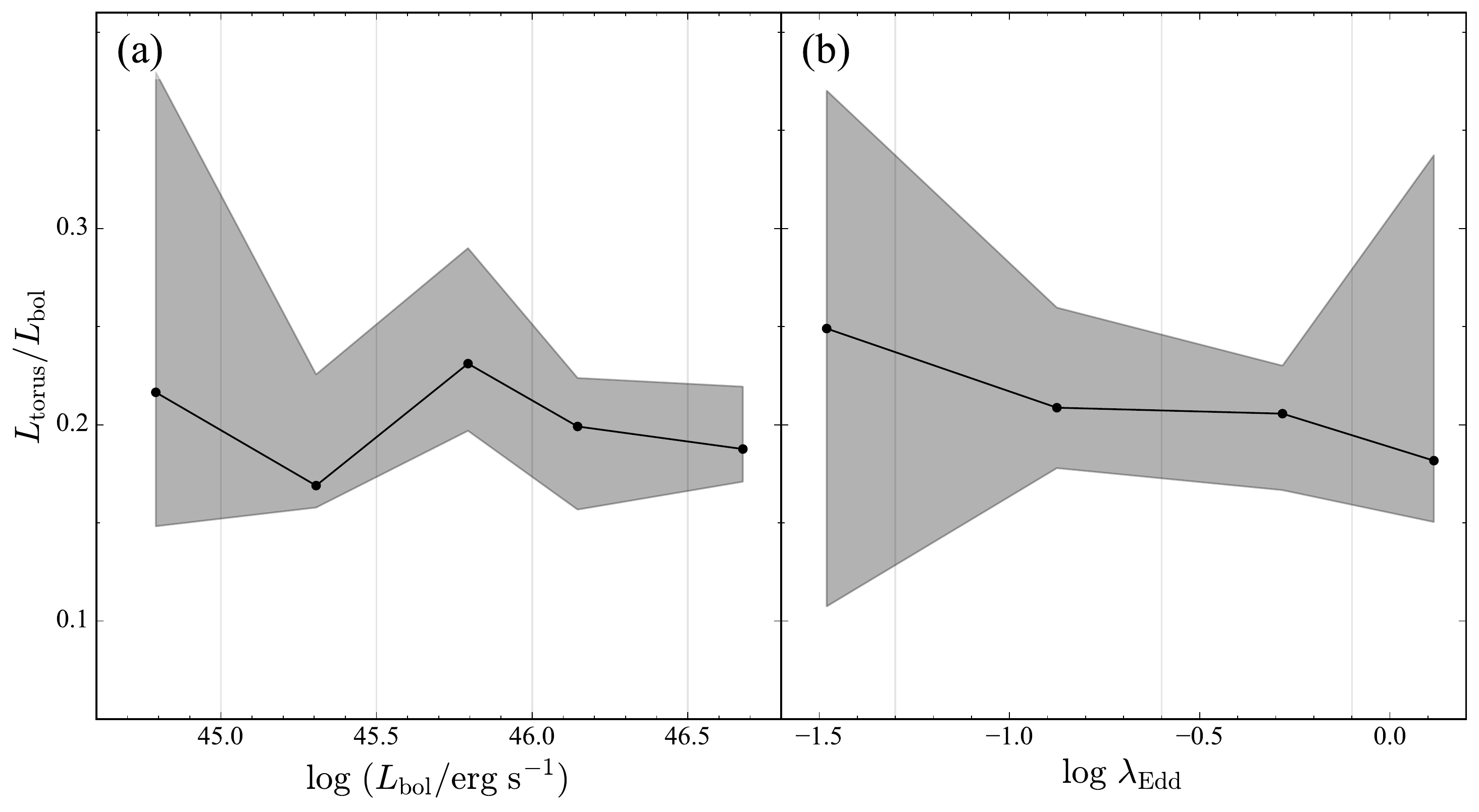}
\caption{The dependence of torus luminosity fraction,
$L_{\mathrm{torus}}/L_{\mathrm{bol}}$, on (a) bolometric luminosity
$L_{\mathrm{bol}}$ and (b) Eddington ratio $\lambda_{\mathrm{Edd}}$.
The sample is divided into bins with the boundaries as indicated by the
vertical lines.  The points are the median values of the bins, with the shaded
area representing the 25th and 75th percentage value of the bins.}
\label{fig:10}
\end{figure*}

\begin{figure}[t]
\centering
\includegraphics[width=0.47\textwidth]{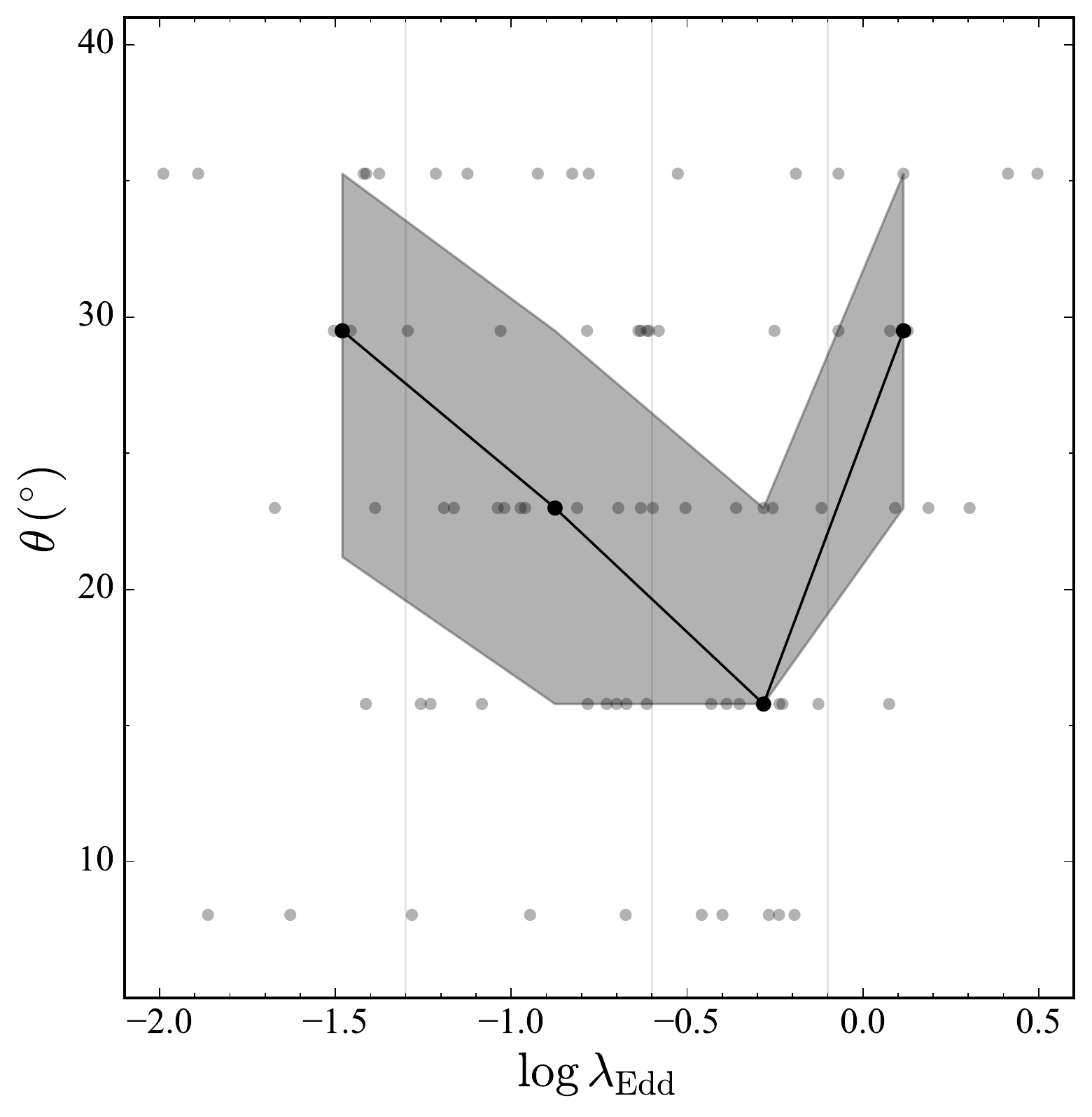}
\caption{
Dependence of the torus half-opening angle $\theta$ on Eddington
ratio $\lambda_{\mathrm{Edd}}$ derived from best-fitting results using the
CAT3D-H-wind torus model. The grey dots in the background represent individual 
objects.  To better visualize any possible trends, we bin the sample into four 
bins of $\lambda_{\mathrm{Edd}}$, whose boundaries are indicated by the 
vertical lines.  The black points are the median values of the four bins, with 
the shaded area representing the 25th and 75th percentage value of their bins.
}
\label{fig:11}
\end{figure}

\begin{figure*}[t]
\centering
\includegraphics[width=\textwidth]{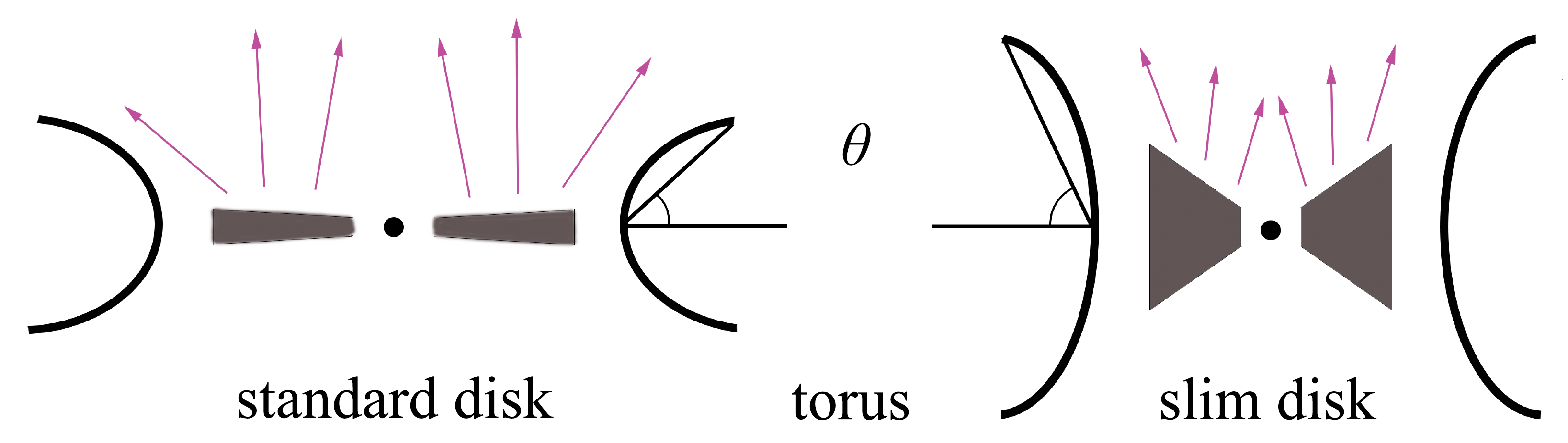}
\caption{Schematic illustration to explain the variation of torus covering 
factor (half-opening angle $\theta$) with Eddington ratio (Figure 
\ref{fig:11}).  For moderate accretion rates, the emission from the standard 
optically thick, geometrically thin accretion disk (left) can suppress the 
vertical height (and hence covering factor) of the torus. At very high or 
super-Eddington accretion rates, the radiation from the optically and 
geometrically thick slim disk (right) is anisotropic and more collimated 
perpendicular to the torus, resulting in large covering factor for the torus.}
\label{fig:12}
\end{figure*}

\section{Discussion}
\label{sec:4}

\subsection{Torus Luminosity and its Contribution to the Total IR Luminosity}
\label{subsec:4.1}

Having established that we can robustly decompose the torus component from the
overall IR SED, we are now able to study the torus luminosity
($L_{\mathrm{torus}}$), in particular its contribution to the total IR
luminosity ($L_{\mathrm{IR}}$), which we define as the sum of all components
from 1 to 1000 $\mu$m.  Appendix \ref{app:C} shows that $L_{\mathrm{torus}}$ 
can be measured robustly in the SED decomposition. 
The correlation between the torus luminosity $L_{\mathrm{torus}}$ and the total 
IR luminosity $L_{\mathrm{IR}}$ is
surprisingly tight (Figure \ref{fig:6}), which indicates that for most of the 
objects the torus contributes a similar fraction of the total energy budget.
The energy fraction is large.  Defining $f_{\mathrm{AGN}}\equiv 
L_{\mathrm{torus}}/ L_{\mathrm{IR}}$, $\langle f_{\mathrm{AGN}} \rangle = 
69.8\% \pm 14.7\%$.  This result implies that the common practice of utilizing 
the integrated IR luminosity to estimate the star formation rates of AGN host 
galaxies may be highly biased, especially for powerful, high-redshift quasars.

\subsection{Correlation between Torus Luminosity and other AGN Bolometric 
Luminosity Tracers}
\label{sec:4.2}

In view of the significant fraction of the IR luminosity radiated by the torus, 
it is of interest to quantify the relationship between the torus luminosity 
and other popularly employed tracers of AGN bolometric luminosity (Figure 
\ref{fig:7}).  A relatively tight (scatter \lax\ $0.3$ dex), essentially linear 
correlation holds between $L_{\mathrm{torus}}$ and the monochromatic 
continuum luminosity at 3000 \AA\ \citep[$L_{3000}$;][]{2004MNRAS.350L..31B}, 
the monochromatic continuum luminosity at 5100 $\mathrm{\AA}$ 
\citep[$L_{5100}$;][]{2006ApJ...641..689V}, and the luminosity of the broad 
H$\beta$ emission line \citep[$L_{\mathrm{H\beta}}$;][]{1992ApJS...80..109B}.  
The strong correlation between $L_{\rm torus}$ and $L_{3000}$ and $L_{5100}$ 
simply reflects the fact that the torus is heated by the ultraviolet/optical continuum 
from the accretion disk.  The correlation between $L_{\rm torus}$ and 
$L_{\mathrm{H\beta}}$, on the other hand, is likely an indirect consequence of 
the more primary relation between the ultraviolet/optical continuum and the broad 
emission lines due to photoionization \citep[e.g.,][]{1980ApJ...241..894Y, 
2005ApJ...630..122G}.  The torus luminosity scales less well (scatter $\sim 
0.5$ dex) with the luminosity of \OIII\ $\lambda$5007 \AA\ 
\citep{1992ApJS...80..109B} from the narrow-line region, presumably because of 
the complex dependence of \OIII\ strength on the intrinsic properties of the 
AGN \citep[e.g., Eddington ratio;][]{2014Natur.513..210S}. 

For convenience, we provide the relations between $L_{\rm torus}$ and 
$L_{3000}$, $L_{5100}$, $L_{\mathrm{H\beta}}$,  and \loiii\ obtained from 
linear least-squares regression\footnote{The observational uncertainties for 
$L_{3000}$, $L_{5100}$, $L_{\mathrm{H\beta}}$,  and \loiii\ are assumed to be 
20\% on a linear scale \citep{2004MNRAS.350L..31B, 2006ApJ...641..689V, 
1992ApJS...80..109B}.  The uncertainties for $L_{\rm torus}$ is from the 
 probability density function (\lax\ 10\% on a linear scale).} method 
\citep{2013MNRAS.432.1709C}, in the form

\begin{equation} \label{eq1}
\log \left(\frac{L_{\rm torus}}{\rm erg\, s^{-1}}\right) = a + b \times 
\left[\log \left(\frac{L_X}{\rm erg\, s^{-1}}\right) - X_0 \right] + \epsilon_{\rm int},
\end{equation}

\noindent
where $\epsilon_{\rm int}$ is the intrinsic scatter. The best-fit parameters 
for Equation \ref{eq1} are given in Table \ref{table:2}.

\begin{deluxetable}{LCCCC}[h]
\tablecaption{Best-fit Linear Regression Parameters for Equation \ref{eq1}
\label{table:2}}
\tablehead{
\nocolhead{$L$} & \colhead{$a$} & \colhead{$b$}& \colhead{$X_0$}& \colhead{$\epsilon_{\rm int}$}
}
\startdata
$L_{3000}$ & $45.120 \pm 0.026$ & $0.859 \pm 0.036$ & $44.95$ & $0.225 \pm 0.022$\\
$L_{5100}$ & $45.109 \pm 0.024$ & $1.022 \pm 0.039$ & $44.74$ & $0.202 \pm 0.019$\\
$L_{\mathrm{H\beta}}$ & $45.171 \pm 0.023$ & $0.866 \pm 0.034$ & $43.03$ & $0.191 \pm 0.019$\\
$L_{\text{[O \textrm{\tiny III}]}}$ & $45.009 \pm 0.043$ & $0.781 \pm 0.064$ & $42.19$ & $0.349 \pm 0.035$\\
\enddata
\end{deluxetable}

\subsection{Properties of Objects with Strong Silicate Emission}
\label{subsec:4.3}

Type 1 quasars commonly exhibit prominent silicate emission features at 
$\sim 9.7 \mu$m and $18 \mu$m (e.g., PG~1259+593; Figure \ref{fig:3}).  
Among our sample of PG quasars, 53 have AGN-dominated mid-IR spectra 
($f_{\mathrm{AGN}}>50\%$) and strong silicate emission [$\nu f_{\nu}$
$(9.7 \mu \mathrm{m})$/$\nu f_{\nu}$$(14 \mu \mathrm{m})>1$].  We examine 
the torus properties of this physically interesting subsample. Figure \ref{fig:8}
shows that objects with strong silicate emission preferentially have fewer 
clouds along the equatorial line-of-sight ($N_0$) and a somewhat
steeper power-law index ($a$) for the radial density profile. A 
Kolmogorov-Smirnov test rejects the null hypothesis that the two subsamples 
are drawn from the same population with a probability of 0.0013 and 0.050, 
respectively.
Taken at face value, this implies that objects with strong silicate emission 
have tori with a more centrally concentrated distribution of clouds.  They 
also have fewer clouds along the line-of-sight to reradiate to longer wavelengths. 
This is consistent with the fact that most silicate-strong objects have very 
strong hot dust emission (Appendix \ref{app:B}). Although the best-fit torus
parameters of individual objects have considerable uncertainty, the overall 
distribution of parameters are relatively robust (Appendix \ref{app:C}). 

\subsection{Anisotropic Emission from the Torus}

Whereas the far-IR emission from the quasar host galaxy is isotropic and 
optically thin, the small-scale torus is optically thick and not spherically 
symmetric, resulting in anisotropic emission in the near-IR and mid-IR 
\citep{1988ApJ...329..702K, 1992ApJ...401...99P}.  Thus, the inclination angle
$i$ of the torus should significantly affect its emission, both in terms of its
luminosity and detailed spectral shape.  When the number of clouds in the wind 
is small compared to that in the torus (e.g., $f_{\mathrm{wd}}=0.15$; Figure 
\ref{fig:9}$a$), similar to the traditional picture of the torus, the SED and 
the flux of the torus change systematically and strongly with $i$, in the 
sense that the average dust temperature and luminosity decrease with 
increasing $i$.  The opposite regime when the number of clouds is much 
larger in the wind than in the torus (e.g., $f_{\mathrm{wd}}=2.25$; Figure 
\ref{fig:9}$b$) presents a very different situation.  The observed torus 
energy first decreases and the SED softens as $i$ increases, and then it 
reverses direction, as a consequence of the wind emission becoming more 
prominent with increasing $i$ (see also Section \ref{sec:3.1}).  In 
cases of extremely large values of $f_{\mathrm{wd}}$, the wind component 
dominates the mid-IR, compensating for or even exceeding the loss of hot 
emission from the torus.

Not all quasars are viewed face-on (Figure \ref{fig:4}).  As a collorary to the 
sensitivity of the torus emission to $i$, the intrinsic, total luminosity 
of the torus itself depends on $i$.  For example, the observed luminosity of a 
quasar with $f_{\mathrm{wd}}=2.25$ and $i=45\degree$ is underestimated by 
nearly a factor of 2 compared to its intrinsic luminosity (at $i=0\degree$; 
Figure \ref{fig:9}$b$).  This implies that the value of $\langle 
f_{\mathrm{AGN}} \rangle \approx 70\%$ (Figure \ref{fig:6}) is most likely a 
lower limit.

\subsection{The Dependence of Torus Opening Angle on Eddington Ratio}
\label{sec:4.5}

The structure of the torus has long been suspected to change with the physical
properties of the AGN.  The most widely discussed concept is that of a receding 
torus \citep{1991MNRAS.252..586L, 2005MNRAS.360..565S, 2007MNRAS.380.1172H}, 
whereby the torus covering factor decreases with increasing $L_{\mathrm{bol}}$,
which manifests itself observationally as an enhanced fraction of type 1 AGNs 
at higher luminosity.  This picture has also enjoyed support from studies that 
parameterize the torus covering factor using the relative luminosity output of 
the torus, finding that $L_{\mathrm{torus}}/L_{\mathrm{bol}}$ decreases with 
increasing AGN luminosity \citep{2007A&A...468..979M, 2008ApJ...679..140T, 
2013ApJ...777...86L}.   More recently, \citet{2017Natur.549..488R}, analyzing a large 
sample of hard X-ray-selected AGNs, proposed that the torus covering factor 
depends primarily not on luminosity but instead on Eddington ratio, 
$\lambda_{\rm Edd}= L_{\mathrm{bol}}/L_{\mathrm{Edd}}$, where 
$L_{\mathrm{Edd}}=1.26 \times 10^{38} (M_{\rm BH}/M_\odot)$.  Radiation 
pressure acting on dust grains expels obscuring material, causing the fraction 
of obscured AGNs---defined in terms of X-ray absorbing column density---to 
decrease with increasing $\lambda_{\rm Edd}$.

Previous studies of the torus covering factor for PG quasars have yielded 
contradictory results.  Analyzing a sample of 64 PG quasars, 
\citet{2005ApJ...619...86C} found no obvious dependence between 
$L_{\rm NIR}/L_{\rm bol}$ and $\lambda_{\mathrm{Edd}}$ or $L_{\rm bol}$.  By 
contrast, \citet{2009ApJ...705..298M}, fitting the SEDs of a subset of 26 PG 
quasars using the CLUMPY torus model (with no additional BB component), 
reported a relatively strong correlation between the model-derived covering 
factor and $L_{\rm bol}$.  We re-examine these trends using our larger 
sample of PG quasars, analyzed with the most updated torus models, using the
BH masses and bolometric luminosities compiled in \citet{2018ApJ...854..158S}. 
Consistent with \citet{2005ApJ...619...86C}, we also see no convincing 
relation between $L_{\mathrm{torus}}/L_{\mathrm{bol}}$ and $L_{\mathrm{bol}}$ 
or $\lambda_{\mathrm{Edd}}$ (Figure~\ref{fig:10}).  As pointed out by 
\citet{2016MNRAS.458.2288S}, the intrinsic anisotropy of the torus makes it 
difficult to estimate the torus covering factor accurately from 
$L_{\mathrm{torus}}/L_{\mathrm{bol}}$. 

We suggest that the best parameter to describe the torus covering factor is 
the torus half-opening angle $\theta$, as it directly relates to the actual
geometry of the torus, independent of inclination. Figure \ref{fig:11} reveals 
an intriguing trend: $\theta$ decreases as the Eddington ratio increases
from $\log \lambda_{\mathrm{Edd}} \approx -2.0$ to $\log \lambda_{\mathrm{Edd}} 
\approx -0.25$; then, $\theta$ rises again as $\log \lambda_{\mathrm{Edd}}$ 
increases from $-0.25$ to $\sim 0.5$.  The large scatter is due to the 
discreteness of the parameter space, the small number of objects in the sample,
as well as to the significant uncertainties of the derived parameters from the 
torus model (Appendix \ref{app:C}).  Despite these limitations, we believe 
that the trends are robust. 

The structure of the accretion disk around a BH changes in response to changes 
in the mass accretion rate.  Three main regimes are commonly recognized: (1) 
an optically thin, geometrically thick, radiative inefficient flow at very low 
accretion rates \citep[$\lambda_{\rm Edd}$ \lax\ 0.01;][and references therein]
{1994ApJ...428L..13N, 2014ARA&A..52..529Y}; (2) an optically thick, geometrically 
thin, standard disk at intermediate accretion rates \citep[0.01 \lax\ $\lambda_{\rm Edd}$ \lax\ 0.1;]
[and references therein]{1973A&A....24..337S, 1998bhad.conf.....K}; 
and (3) an optically thick, geometrically thick, slim disk at very high 
accretion rates \citep[$\lambda_{\rm Edd}$ \gax\ 0.3;][and references therein]
{1978MNRAS.184...53B, 1988ApJ...332..646A}.  The relationship between the torus 
opening angle and Eddington ratio may arise from the interplay between the 
illumination pattern of the central accretion disk and its surrounding torus.  
Consider the schematic sketched in Figure \ref{fig:12}.  When 
$\lambda_{\rm Edd}$ is low enough for the accretion flow to be radiatively 
inefficient (not illustrated), the deficit of ultraviolet photons in its SED \citep{1999ApJ...516..672H, 
2008ARA&A..46..475H} implies that a large fraction of the dusty gas \citep[if present; see]
[]{She2018} cannot be evacuated by radiation pressure, resulting in a 
large torus covering factor.  With increasing $\lambda_{\rm Edd}$ the 
accretion disk enters the standard regime, and its large ultraviolet output can 
efficiently clear away the obscuring material, leading to a systematic 
decrease in $\theta$.  Finally, when $\lambda_{\rm Edd}$ crosses above the 
threshold for a slim disk, its vertically thick inner funnel results in 
significant anisotropy of its ionizing radiation field \citep{1988ApJ...332..646A, 
2014ApJ...797...65W}, which again leads to an increase in $\theta$.  Note that a 
natural corollary of this model is that highly accreting (super-Eddington) 
AGNs should contain a larger fraction of type 2 sources.

Our schematic picture is qualitatively consistent with but expands upon the 
scenario proposed by \citet{2017Natur.549..488R}, primarily by extending the 
dynamic range to $\lambda_{\rm Edd} > 1$, since the PG quasar sample contains a 
sizable fraction of highly accreting sources.  By the same token, Ricci et 
al.'s sample, selected by hard X-rays from {\it Swift}/BAT observations, 
extends $\lambda_{\rm Edd}$ to significantly lower values than our sample.  
Hence our two studies are highly complementary.

\section{Conclusions}
\label{sec:conclusion}

We apply a newly developed Bayesian MCMC method to study the IR ($1-500\,\mu$m) 
SEDs of a large, well-defined sample of low-redshift ($z < 0.5$) Palomar-Green 
quasars.  Our primary motivation is to quantify the properties of the 
AGN-heated dust, by decomposing the SEDs using a combination of physically 
motivated emission components for the stars, torus, and large-scale dust 
component of the host galaxy.  Our extensive tests of a suite of theoretical 
templates indicate that the majority of the quasar SEDs can be best fitted with the 
torus models of \citet[][CAT3D]{2017ApJ...838L..20H} that properly account 
for the different sublimation temperatures of silicate and graphite grains and 
consider a polar wind component.  

Our main conclusions are the following:

\begin{enumerate}

\item The luminosity of the torus correlates tightly with the total IR 
($1-1000\,\mu$m) luminosity.  On average, 
$\langle L_{\mathrm{torus}}/L_{\mathrm{IR}}\rangle \approx 70\%$.  
Star formation rates of quasar host galaxies estimated using $L_{\mathrm{IR}}$ 
will be significantly overestimated if the contribution from the torus is not 
properly taken into account.

\item The luminosity of the torus correlates tightly (scatter $<0.3$ dex) with 
the luminosity of the ultraviolet/optical continuum and the broad $\rm{H\beta}$ emission line, indicating a close link between the central ionization source 
and re-radiation by the torus.  

\item The majority of the torus inclination angles lie in the range \lax\ 
$45\degree$, consistent with expectations for type~1 (broad-line) AGNs.

\item Most PG quasars (53/76) show both strong hot dust emission 
and silicate features, which can be used to differentiate dust torus models.

\item The torus covering factor, as reflected in the torus half-opening angle 
$\theta$, decreases with increasing Eddington ratio until $\lambda_{\rm Edd} 
\approx 0.5$, above which $\theta$ rises again.  We suggest that these 
trends can be explained by the pattern of the radiation field impinging upon 
the torus from the accretion disk, which transitions from a standard thin disk 
to a slim disk at the highest accretion rates.

\end{enumerate}

\begin{acknowledgements}
We thank the referee, M. Kishimoto, for his stimulating comments and suggestions 
that helped to improve the quality and presentation of our paper.
This work was supported by the National Key R\&D Program of China
(2016YFA0400702) and the National Science Foundation of China (11473002,
11721303, 11403072).
\end{acknowledgements}

\appendix

\section{Influence of Different Torus Models on the Derived Dust Masses}
\label{app:A}

\begin{figure}[h]
\centering
\includegraphics[width=0.6\textwidth]{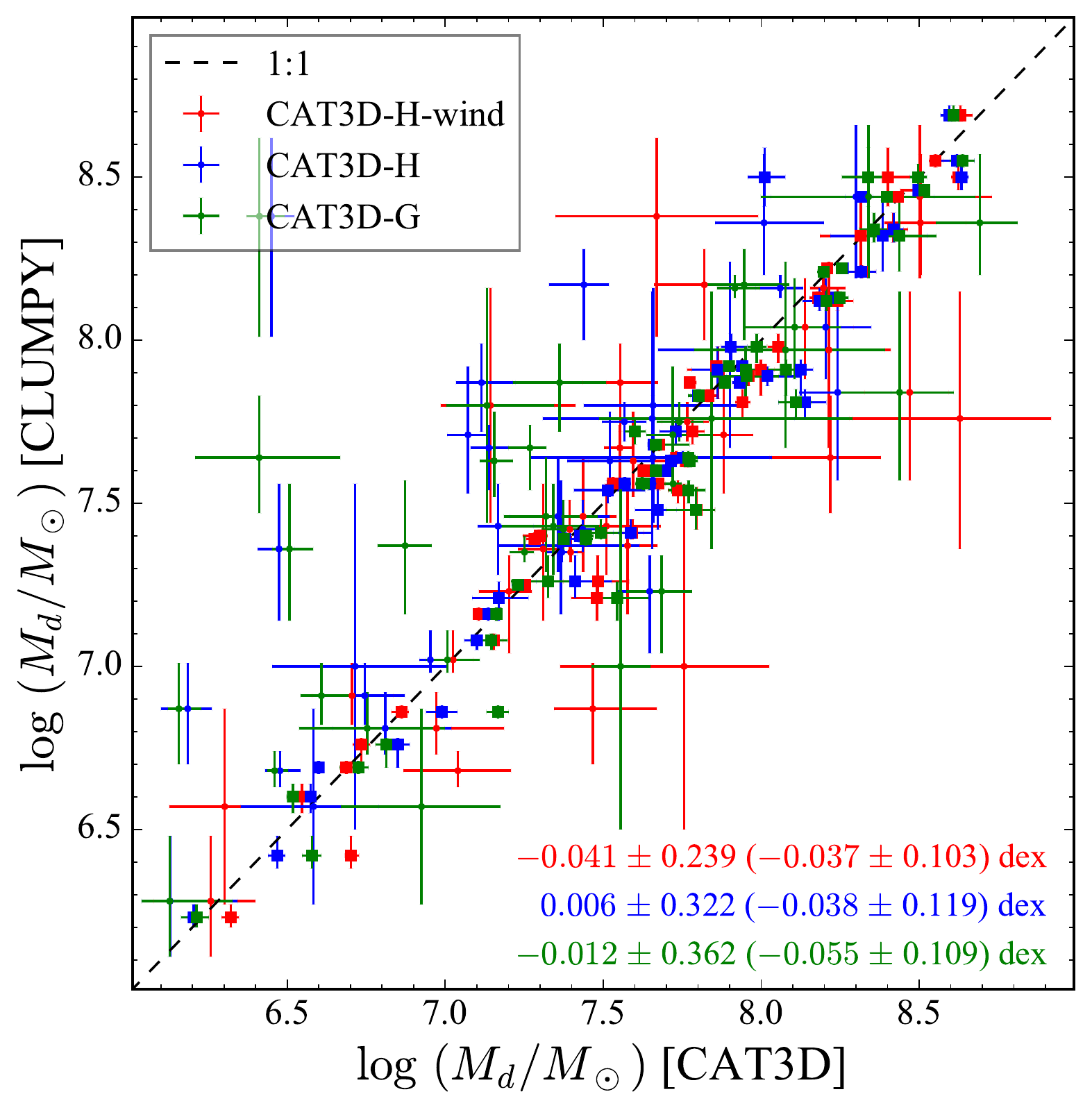}
\caption{Comparison of cold dust masses for the host galaxy derived with 
different torus models. Colors are the same as Figure \ref{fig:4}.  Large 
points represent objects with SEDs whose far-IR peak is well-constrained by the 
{\it Herschel}\ data; smaller points do not have well-constrained far-IR peaks.
The residuals and standard deviations are given in the lower-right corner; 
values in parentheses are for the objects with SEDs with well-constrained 
far-IR peaks.}
\label{fig:13}
\end{figure}

Our fitting code derives, as a by-product, the interstellar cold dust 
mass $M_d$ from the DL07 model. Figure \ref{fig:13} shows the effect on $M_d$ 
of choosing different torus models for the SED fitting, using, as reference,
the CLUMPY model. As long as the far-IR peak of the SED is well constrained, no 
significant difference is found on $M_d$.  Any systematic deviations are at 
the level of $\lesssim 0.05$ dex, with standard deviations of $\sim 0.1-0.3$ dex. 
This is consistent with the results of \citet{2018ApJ...854..158S}, who compared 
the effect on $M_d$ from the use of the CLUMPY torus model and another torus 
model by \citet{2017ApJS..228....6X}.

\begin{figure}[h]
\centering
\includegraphics[width=0.6\textwidth]{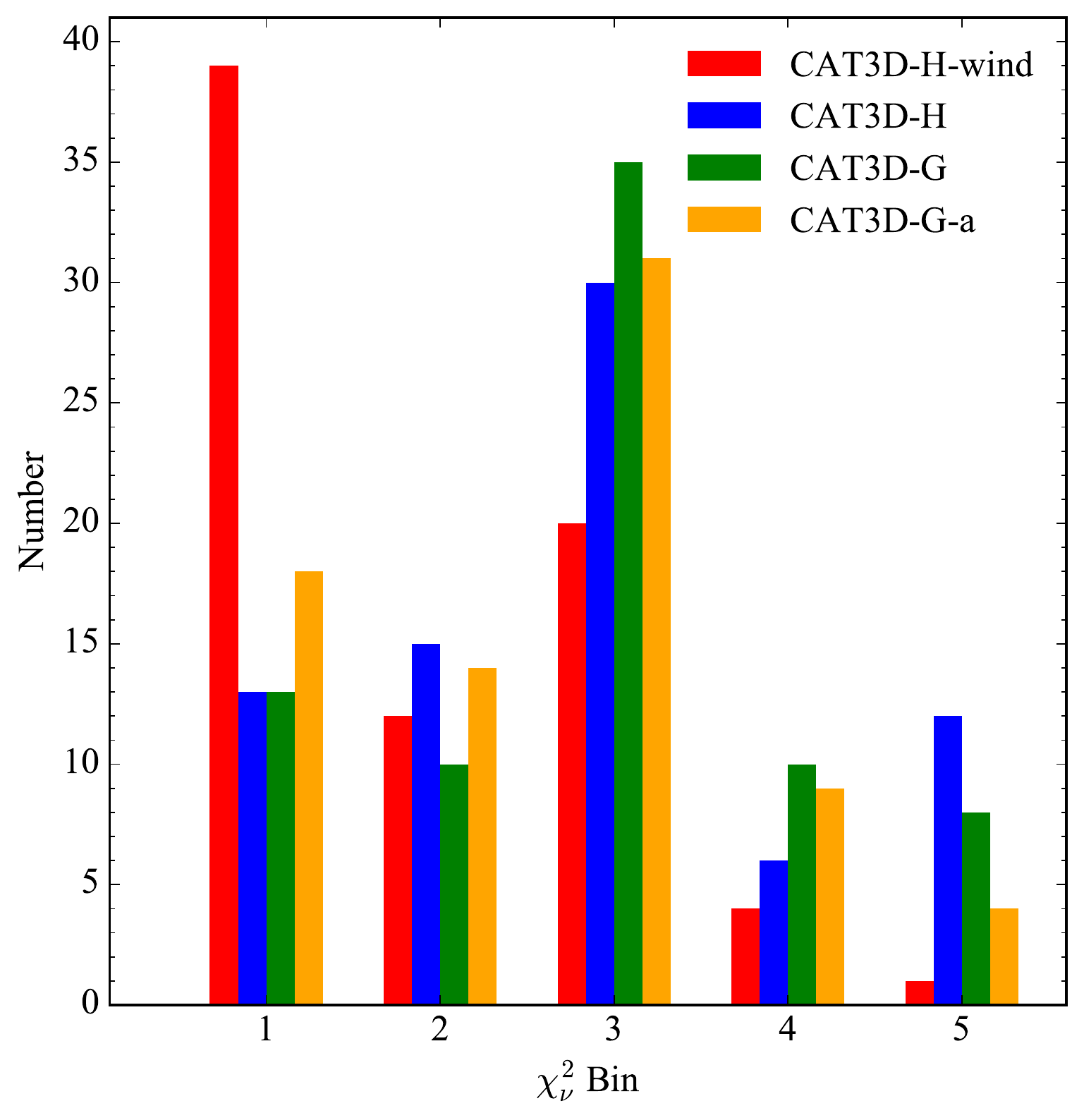}
\caption{
Histogram of $\chi_{\nu}^{2}$ of fitting results for the CAT3D-H-wind (red), 
CAT3D-H (blue), CAT3D-G (green), and CAT3D-G-a (orange) torus models. Bins for $\chi_{\nu}^{2}$ are (1) 0--50, (2) 50--100, (3) 100--500, (4) 500--1000, and (5) $>$1000.}
\label{fig:14}
\end{figure}

\begin{figure}[h]
\centering
\includegraphics[width=0.8\textwidth]{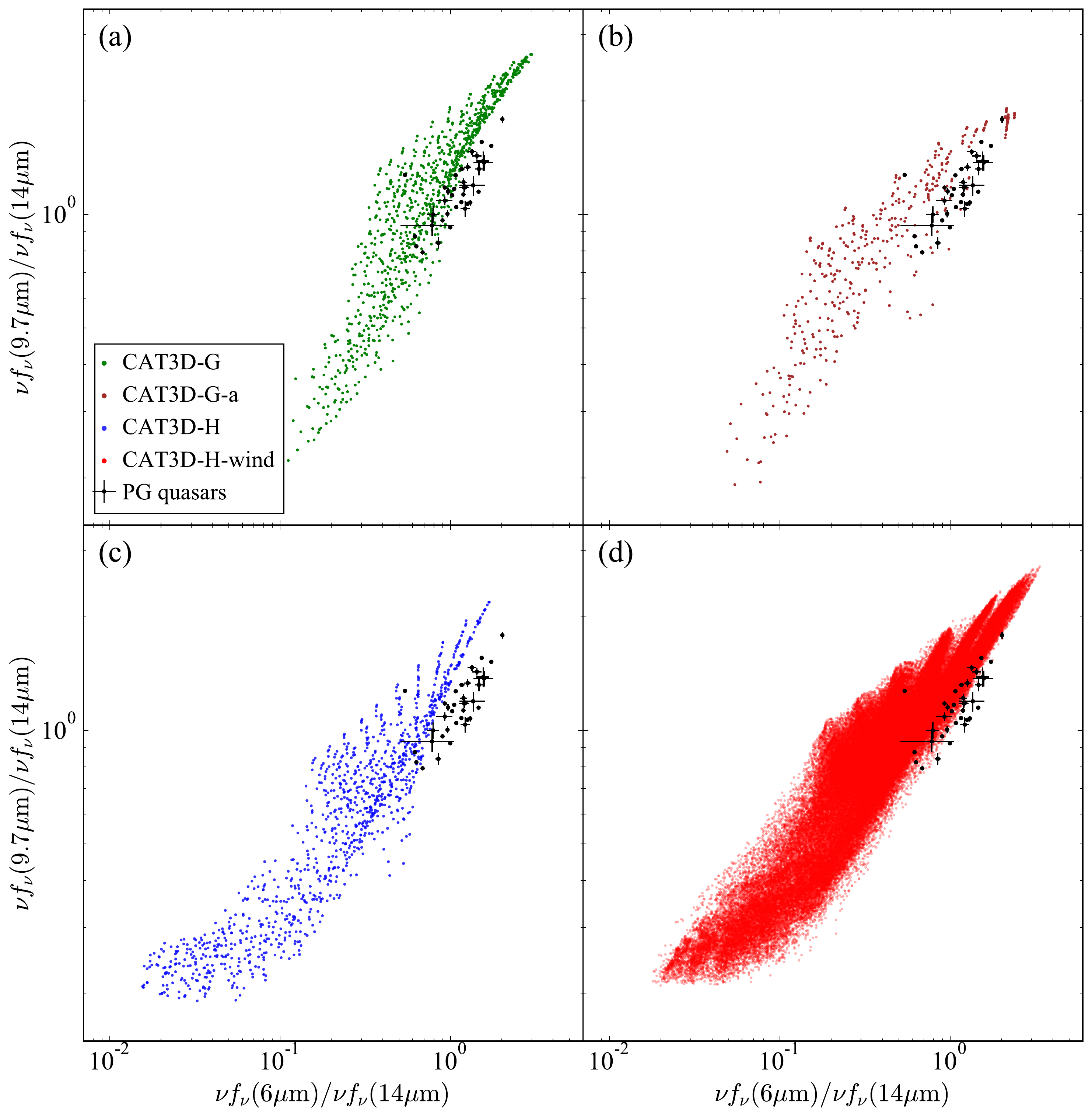}
\caption{
Comparison of different model coverages of strength of hot dust [$\nu f_{\nu} (6 \mu \mathrm{m})/\nu f_{\nu}(14 \mu \mathrm{m})$] and silicates [$\nu f_{\nu}(9.7 \mu \mathrm{m})/\nu f_{\nu}(14 \mu \mathrm{m})$] for our sample. Model coverage of CAT3D-G, CAT3D-G-a, CAT3D-H, and CAT3D-H-wind are shown in panels (a) to (d), respectively.  ``PG quasars'' represents the values derived from {\it Spitzer}/IRS spectrum for objects with AGN fraction ($f_{\rm AGN}$) above the median value of the whole sample.
}
\label{fig:15}
\end{figure}

\section{Goodness-of-Fit of Different Torus Models}
\label{app:B}

We use the reduced $\chi_{\nu}^2$ (${\chi^2}$ per degree of freedom) to 
assess quantitatively the relative goodness-of-fit of the different 
torus models. As the {\it Spitzer}/IRS spectrum covers the spectral range that 
maximally constrains the dust torus, we restrict the $\chi_{\nu}^2$ calculation
to the wavelength region $\sim 5-38\,\micron$ (in the observed frame).
For cases without strong silicate features and hot dust emission (e.g., 
PG~0934+013; Figure \ref{fig:2}), all the models provide similarly acceptable 
fits. 
A large fraction of the objects, however, resemble PG~1259+593 (Figure \ref{fig:3}), for 
which the CAT3D-H-wind model clearly exhibits a much lower $\chi_{\nu}^2$ 
than the other models\footnote{The notable exception is the CLUMPY+BB model, 
which, despite its flexibility, is disfavored because of the ad hoc nature 
of the BB component (Section \ref{sec:3.1}).}.  To compare the different models 
quantitatively, we group the $\chi_{\nu}^2$ values into five bins: 0--50, 
50--100, 100--500, 500--1000, and $>$1000. Figure \ref{fig:14} shows that more than 
half of the objects fit using CAT3D-H-wind have $\chi_{\nu}^2 < 50$.  Figure 
\ref{fig:2} illustrates that $\chi_{\nu}^2 \approx 50$ already signifies a 
very good fit.  By contrast, the other three CAT3D models that lack a wind 
component have $\chi_{\nu}^2$ distributions that peak in the range 
$\sim 100-500$, nearly an order of magnitude larger than the wind model. 
In extreme cases such as PG~1259+593 (Figure \ref{fig:3}), the wind model 
provides exceptionally good fits with $\chi_{\nu}^2 \approx 10$, 
unquestionably superior to the non-wind models.  Thus, we conclude, 
based on the sample as a whole and on certain individual cases, that 
the CAT3D torus models that incorporate a wind component best match the 
mid-IR SEDs of our sample of quasars.

What aspects of the mid-IR SED actually distinguish the wind model from the 
others?  Two prominent features stand out in our quasar sample: hot dust 
continuum emission at $\sim$5 $\mu$m and strong silicate emission at $\sim$9.7 
and 18 $\mu$m.  
We use the flux ratio of the continuum emission at 6 $\mu$m\footnote{In order to 
be covered by {\it Spitzer}/IRS spectrum and avoid possible PAH emission.} and 
14 $\mu$m 
to indicate the relative strength of the hot dust emission, and the flux ratio of the 
9.7 $\mu$m silicate emission to that of the 14 $\mu$m continuum emission to 
gauge the relative strength of the silicate features. 
Figure \ref{fig:15} shows these flux ratios for the 38 PG quasars with $f_{\rm AGN}
>69.8\%$ (median value of the whole sample) derived from the observed spectrum.  
Our quasar sample exhibits hot dust emission and silicate emission of comparable 
relative strength: the median value of $\nu f_\nu(6\micron)/\nu f_\nu(14\micron) 
\approx 1.18$ and $\nu f_\nu(9.7\micron)/\nu f_\nu(14\micron) \approx 1.16$. 
The two ratios show a relatively strong correlation with a Pearson's correlation 
coefficient of $r \approx 0.8$ and a $p$-value of $\sim10^{-9}$.
For comparison, we overlay the flux ratios computed from the theoretical template 
spectra for the four versions of the CAT3D models. 
The non-wind models (Figure \ref{fig:15}, panels (a)--(c)) cover little of the observed 
parameter space of the PG quasars, whereas the CAT3D-H-wind templates cover most 
of it (panel (d)). Objects whose flux ratios are not covered by the CAT3D-H-wind 
templates have relatively worse fits. 
Two non-wind models (CAT3D-G and CAT3D-H), would always predict much stronger 
silicate emission for a given hot dust strength. For torus models that only have a 
toroidal structure, the spatial distributions of the graphite and silicate dust are 
coupled, except for temperatures $> 1200$ K. The observed strong hot dust emission 
requires the model to have a highly concentrated dust distribution.  As the clump 
distribution is described by a power law, the silicates will have the same centrally 
concentrated distribution as the graphite and will be heated to high temperature.  
The introduction of an extra wind component, however, contributes more hot graphite 
emission without boosting the strength of the silicate emission \citep{2017ApJ...838L..20H}. 
The poor coverage of CAT3D-G-a model is possibly due to the limited number of templates.

\begin{figure}[!ht]
\centering
\includegraphics[width=0.8\textwidth]{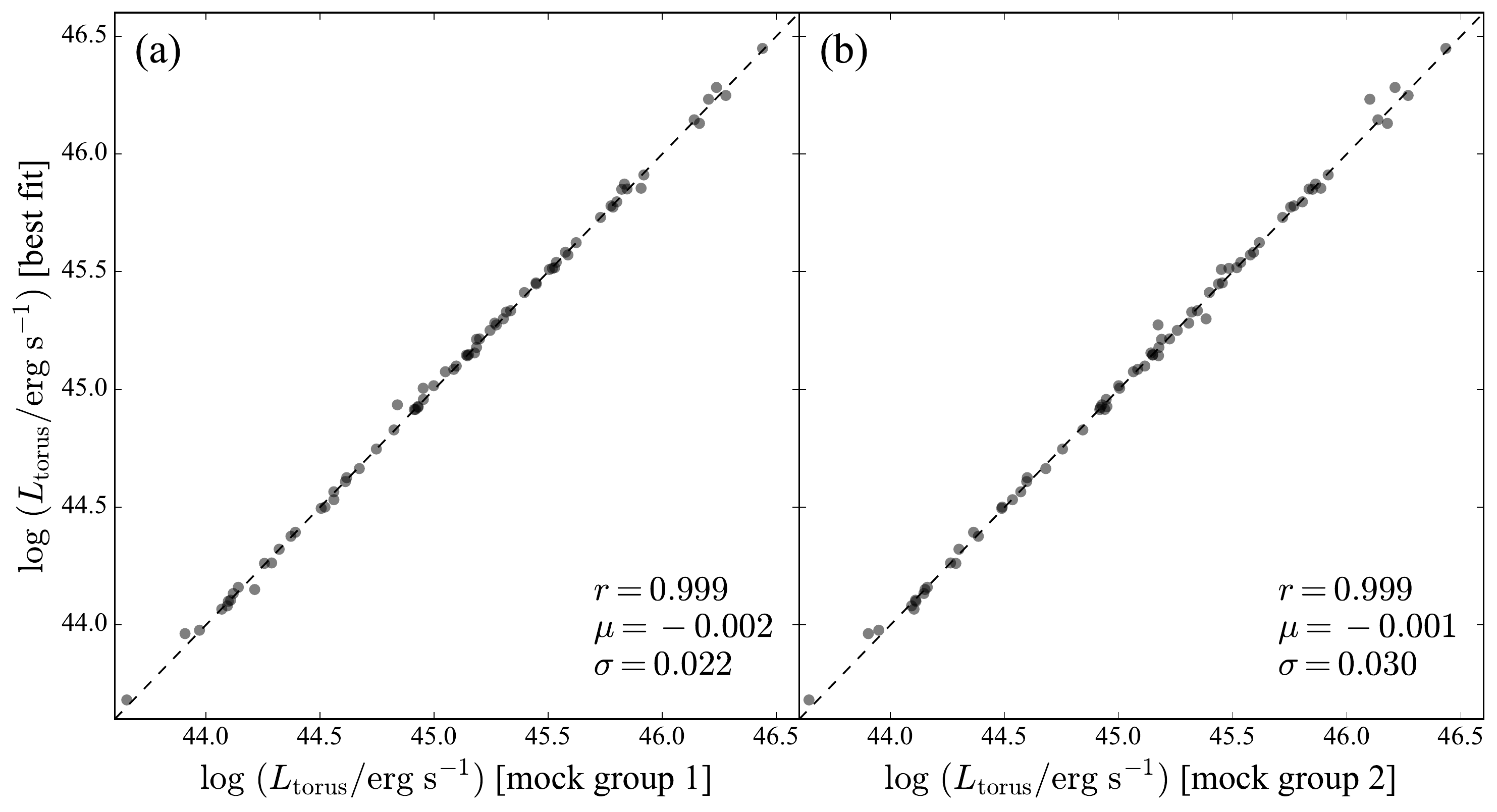}
\caption{Comparison of torus luminosity $L_{\mathrm{torus}}$ from the best fit
with that from (a) mock group 1 and (b) mock group 2. The legend gives the 
Pearson's correlation coefficient ($r$), median ($\mu$), and standard deviation ($\sigma$).}
\label{fig:16}
\end{figure}

\begin{figure}[!ht]
\centering
\includegraphics[width=0.8\textwidth]{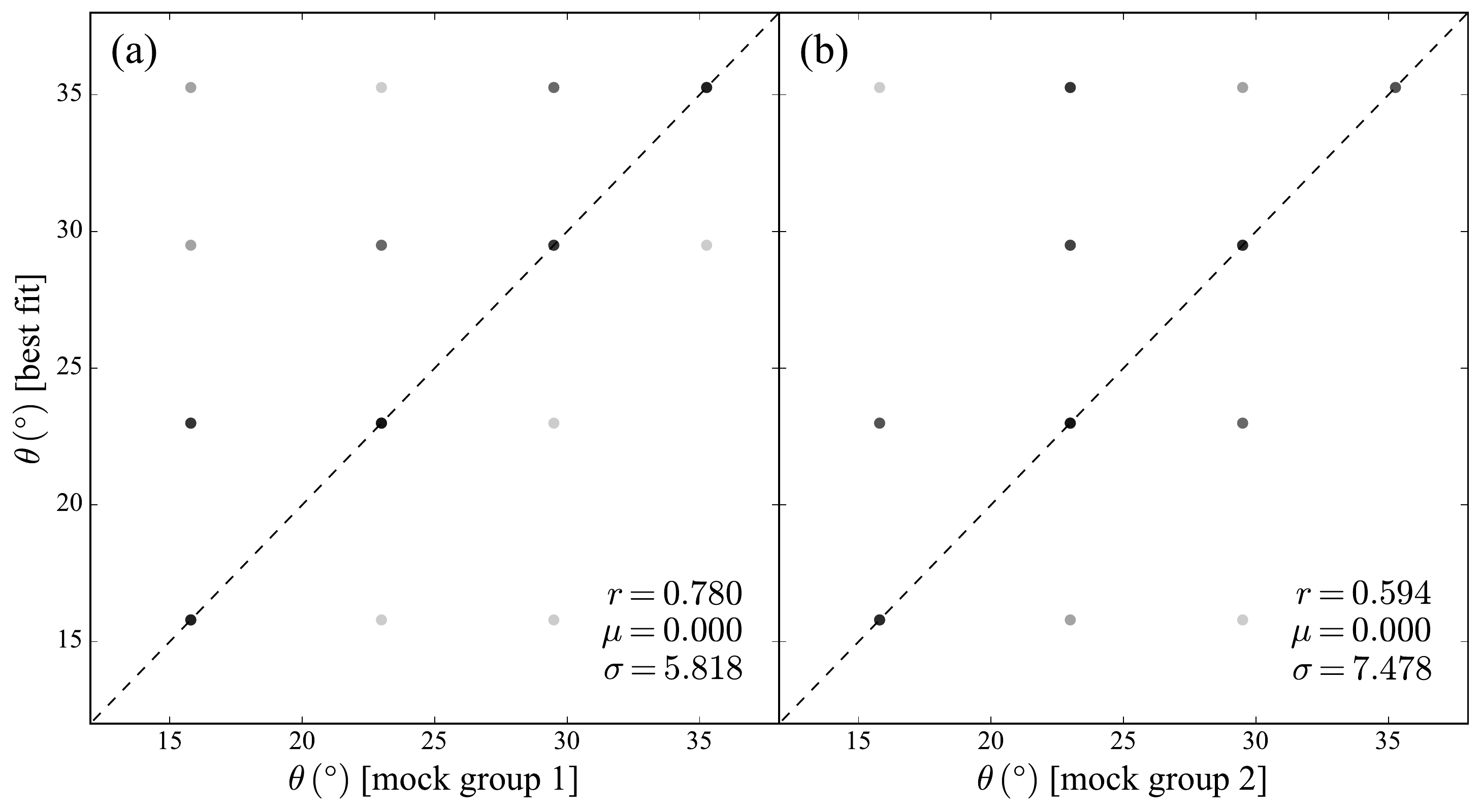}
\caption{Same as Figure \ref{fig:16}, but for half-opening angle $\theta$.}
\label{fig:17}
\end{figure}

\section{Sensitivity of Parameters to the Data}
\label{app:C}

We use mock data to test the robustness of our fitting results and explore the 
sensitivity of the parameters to the data, taking comprehensive uncertainties into account. 
Following the methodology of \citet{2018ApJ...854..158S}, 
we generate two groups of mock SEDs using the best-fit CAT3D-H-wind models of 
the real PG quasar SEDs.  The two groups of mock SEDs are generated as 
follows\footnote{We exclude three objects lacking complete measurements in all 
six {\it Herschel}\ bands.}: 

\begin{enumerate}
\item We use the best-fit parameters of each quasar to generate the SED 
model.  The mock data are calculated from the SED model, perturbed according 
to their uncertainties from the observed data, assuming a Gaussian distribution 
with standard deviation equal to their uncertainties.

\item We use the values from (1) but add systematic uncertainty to the 
photometry and spectrum to account for calibration uncertainties, assuming a 
Gaussian distribution with different standard deviations: 5\% for 
{\it Spitzer}/IRS, 3\% for 2MASS and {\it WISE}, and 5\% for 
{\it Herschel}/PACS and SPIRE. The systematic uncertainty for each band from 
the same data set is the same (e.g., the uncertainties for the $J$, $H$, and 
$K_s$ bands of 2MASS are the same). Then, we substitute the values for the {\it Herschel} 
bands with upper limits if these bands are not detected in the real data.

\end{enumerate}

We fit the two groups of mock SEDs using the CAT3D-H-wind model.  The fit 
results are globally very good.  For group~1, the best-fit models are always 
indistinguishable from the data; for group~2, although additional systematic 
perturbation is applied, the best fits still match the data very well.  
The input and best-fit torus models usually overlap each other closely,
especially, at $\sim 5-20\,\micron$, where the torus component
dominates the entire model.  However, for some cases the input and best-fit
torus models start to deviate at $\lambda >20\,\micron$,
mainly because the DL07 component starts to couple with the torus
component.  The large discreteness of the DL07 parameters makes the
fitting challenging to find the true values \citep{2018ApJ...854..158S}.
When comparing the input and best-fit parameters, we find that the 
exact input parameters are not easily reproduced in the fitting, likely 
because of the difficulty of distinguishing between subtle variations of the 
torus models, as well as the degeneracy between the torus and the DL07 model.  
The torus model is especially degenerate for the
wind component, as templates generated from different configurations of model
parameters produce only subtle differences in the resulting SEDs.  Thus,
caution should be exercised in interpreting the best-fit parameters of the
torus model for individual objects, especially parameters associated with the
wind component.  In contrast to the uncertainties of the wind component, the
integrated luminosity of the torus ($L_{\rm torus}$) can be recovered very
reliably ($\sim 0.02-0.03$ dex; Figure \ref{fig:16}).  The torus half-opening angle
($\theta$) also suffers from no systematic bias, although the scatter is
substantial ($\sim 6\degree-7\degree$; Figure \ref{fig:17}). Thus, the overall
statistical trend between $\theta$ and $\lambda_{\mathrm{Edd}}$ 
(Figure \ref{fig:11}) should be robust, even if the large observed scatter may 
be due to uncertainty in individual measurements.

\end{document}